# NOISE IN MESOSCOPIC PHYSICS

T. Martin

*Centre de Physique Théorique et Université de la Méditerranée*
*case 907, 13288 Marseille, France*

Photo: width 7.5cm height 11cm

# Contents

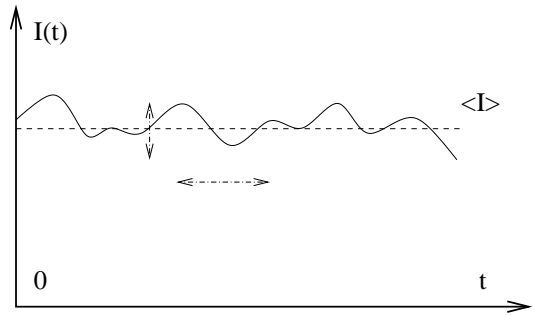

Fig. 1. The current as a function of time undergoes fluctuation around an average value, the height of the "waves" and their frequency are characterized by the noise.

## 1. Introduction

When one applies a constant bias voltage to a conductor, a stationary current is typically established. However, if one is to carefully analyze this current, one discovers that it has a time dependence: there are some fluctuations around the average value (Fig. 1). One of the ways to characterize these fluctuations is to compute the current-current correlation function and to calculate its Fourier transform: the noise. The information thus obtained characterizes the amplitude of the deviations with respect to the average value, as well as the frequency of occurrence of such fluctuations.

Within the last decade and a half, there has been a resurgence of interest in the study of noise in Mesoscopic devices, both experimentally and theoretically. In a general sense, noise occurs because the transport of electrons has a stochastic nature. Noise can arise in classical and quantum transport as well, but here we will deal mostly with quantum noise. Noise in solid state devices can have different origins: there is $1/f$ noise [1] which is believed to arise from fluctuations in the resistance of the sample due to the motion of impurities [2] and background charges. This course does not deal at all on $1/f$ noise. The noise I am about to describe considers the device/conductor as having no inelastic effects: a probability of transmission, or a hopping matrix element characterizes it for instance. Typically in experiments, one cannot dissociate $1/f$ noise from this "other" noise contribution. $1/f$ noise can obviously not be avoided in low frequency measurements, so experimentalist typically perform measurements in the kHz to MHz range, sometimes higher.

If the sample considered is small enough that dephasing and inelastic effects can be neglected, equilibrium (thermal) noise and excess noise can be completely described in terms of the elastic scattering properties of the sample. This is the regime of mesoscopic physics, which is now described in several textbooks [3]. As mentioned above, noise arises as a consequence of random processes governing the transport of electrons. Here, there are two sources randomness: first, electrons incident on the sample occupy a given energy state with a probability given by the Fermi–Dirac distribution function. Secondly, electrons can be transmitted across the sample or reflected in the same reservoir where they came from with a probability given by the quantum mechanical transmission/ reflection coefficients.

Equilibrium noise refers to the case where no bias voltage is applied between the leads connected to the sample, where thermal agitation alone allows the electrons close to the Fermi level to tunnel trough the sample [4]. Equilibrium noise is related to the conductance of the sample via the Johnson–Nyquist formula [5]. In the presence of a bias, in the classical regime, we expect to recover the full shot noise, a noise proportional to the average current. Shot noise was predicted by Schottky [6] and was observed in vacuum diodes [7].

If the sample is to be described quantum mechanically, a calculation of the noise should include statistical effects such as the Pauli principle: an electron which is successfully transmitted cannot occupy the same state as another electron incident from the opposite side, which is reflected by the potential barrier. The importance of the statistics of the charge carriers is a novelty in Mesoscopic physics. After all, many experiments in Mesoscopic physics – with electrons as carriers – have a direct optical analog if we interchange the carriers with photons. The conductance steps experiment [8] which shows the transverse quantization of the electron wave function is an example: this experiment has been successfully completed with photons [9]. The measurement of universal conductance fluctuations in Mesoscopic wires and rings [10, 11] also has an analog when one shines a laser on white paint or cold atoms as one studies the retrodiffusion peak or the speckle pattern which is generated [12]. In contrast to these examples, a noise measurement makes a distinction between fermions and bosons.

Many approaches have been proposed to calculate noise. Some are quasi–classical. Others use a formulation of non-equilibrium thermodynamics which is based on the concept of reservoirs, introduced for the conductance formula [13]. Here, we shall begin with an intuitive picture [14–16], where the current passing through the device is a superposition of pulses, or electrons wave packets, which can be transmitted or reflected. We will then proceed with scattering theory, and conclude with calculations of noise with the Keldysh formalism for a strongly correlated system.

The scattering approach based on operator averages [17–19] will be described in detail, because it allows to describe systematically more complex situations where the sample is connected to several leads. It also allows a generalization to finite frequency noise and to conductors which have an interface with a superconductor. Superconductors will be studied not only for their own sake, but also because they provide a natural source of entangled electrons. Noise correlation measurements can in principle be used to test quantum mechanical non-locality. Starting from a microscopic Hamiltonian, one can use non-equilibrium formulations of field theory and thermodynamics [20] to compute the current, as well as the noise [21–23]. There are few systems in mesoscopic physics where the role of electronic interactions on the noise can be probed easily. Coulomb interaction give most of us a considerable headache, or considerable excitement. One dimensional systems are special, in the sense that they can be handled somewhat exactly. In the tunneling regime only, I will discuss the noise properties of Luttinger liquids, using the illustration of edge channels in the fractional quantum Hall effect. Even the lowest order tunneling calculation brings out an interesting result: the identification of anomalous, fractional charges via the Schottky formula.

Noise also enters in dephasing and decoherence processes, for instance, when a quantum dot with a sharp level is coupled electrostatically to the electrons which transit in a nearby wire [24].

Throughout the course, we will consider conductors which are connected to several terminal because one can build fermionic analogs of optical devices for photons [25].

Finally, I mention that there are some excellent publications providing reviews on noise. One book on fluctuations in solids is available [26], and describes many aspects of the noise – such as the Langevin approach – which I will not discuss here. Another is the very complete review article of Y. Blanter and M. Büttiker [27], which was used here in some sections. To some extent, this course will be complementary to these materials because it will present the Keldysh approach to noise in Luttinger liquids, and because it will discuss in what way noise correlations can be probed to discuss the issue of entanglement.

## 2. Poissonian noise

Walter Schottky pioneered in 1918 the field by calculating the noise of a source of particles which are emitted in an independent manner [6]. Let $\tau$ correspond to the mean time which separates two tunneling events. In quantum and classical transport, we are dealing with situations where many particle are transmitted from one lead to another. Consider now the probability $P_N(t)$ for having $N$ tunneling events during time $t$. This follows the probability law:

$$P_N(t) = \frac{t^N}{\tau^N \, N!} e^{-t/\tau} \ . \tag{2.1}$$

This can be proved as follows. The probability of $N$ tunneling events can be expressed in terms of the probability to have $N-1$ tunneling events:

$$P_N(t+dt) = P_{N-1}(t)\frac{dt}{\tau} + P_N(t)\left(1 - \frac{dt}{\tau}\right) \tag{2.2}$$

which leads (after multiplying by $\exp(t/\tau)$) to

$$\frac{d}{d(t/\tau)} \Pi_N = \Pi_{N-1} \ , \tag{2.3}$$

where $\Pi_N \equiv P_N \exp(t/\tau)$. The solution by induction is clearly

$$\Pi_N = (t/\tau)^N \Pi_0 / N! \tag{2.4}$$

Because the solution for $P_0$ is given by:

$$\frac{dP_0}{dt} = -\frac{1}{\tau} P_0 \ , \qquad P_0 = \exp(-t/\tau) \ . \tag{2.5}$$

One finally obtains the above result for $P_N(t)$, which can be expressed in turn as a function of the mean number of particles transmitted in $t$.

$$P_N(t) = \frac{\langle N \rangle^N}{N!} e^{-\langle N \rangle} \ . \tag{2.6}$$

To show this, one considers the characteristic function of the distribution:

$$\phi = \sum_N \frac{e^{i\lambda N}(t/\tau)^N}{N!} e^{-t/\tau} = e^{(t/\tau)(e^{i\lambda}-1)} \ . \tag{2.7}$$

The characteristic function gives the average number of particle transmitted when differentiating it respect to $\lambda$ once (and setting $\lambda = 0$), and the variance when differentiating with $\lambda$ twice. We thus get:

$$\langle N \rangle = t/\tau, \qquad \langle N^2 \rangle - \langle N \rangle^2 = \langle N \rangle \ . \tag{2.8}$$

This has a fundamental consequence on the noise characteristic of a tunnel junction. The current is given by

$$\langle I \rangle = e\langle N \rangle / t = e/\tau \ . \tag{2.9}$$

The noise is proportional to the variance of the number of particles transmitted. Here we give the final result for the spectral density of noise:

$$S = 2e^2(\langle N^2 \rangle - \langle N \rangle^2)/t = 2e\langle I \rangle \,. \tag{2.10}$$

This formula has the benefit that it applies to any tunneling situation. It applies to electrons tunneling between two metallic electrodes, but also to "strange" quasi-particles of the quantum Hall effect tunneling between two edge states. Below we shall illustrate this fact in discussing the detection of the quasiparticle charge. At the same time this formula will serve as a point of comparison with the quantum noise derivations which apply to mesoscopic conductors.

## 3. The wave packet approach

We consider first a one dimensional sample connected to a source and a drain. The results presented where first described in [17, 28] with other methods. The quantity we wish to calculate is the time correlation in the current:

$$C(t) = \frac{1}{T} \int dt' \, \langle I(t')I(t+t') \rangle \,. \tag{3.1}$$

The spectral density of noise $S(\omega)$ is related to the above quantity by a simple Fourier transform. The measurement frequencies which we consider here are low enough compared to the inverse of the time associated with the transfer of an electron from source to drain [29] and allow to neglect the self inductance of the sample. Using the Fourier representation for the current, this yields:

$$S(\omega) = \lim_{T \to \infty} \frac{2}{T} \langle |I(\omega)|^2 \rangle \,, \tag{3.2}$$

where the angular brackets denote some kind of average over electrons occupation factors. The wave packet approach views the current passing across the sample as a superposition of clocked pulses [16]: $I(t) = \sum_n j(t-n\tau)g_n$ . In this expression, $j(t)$ is the current associated with a given pulse and $g_n$ is an occupation factor which takes a value 1 if an electron has been transferred from the left hand side to the right hand side of the sample, $-1$ if the electron was transferred from right to left, and $0$ when no electron is transferred at all. The quantum mechanics necessary to calculate the noise is hidden in $g_n$. The wave packets representing the electrons are separated in time, but can overlap in space. An example of wave packet construction can be obtained if we consider states limited to a small energy interval $\Delta E$ [16]: choosing $\tau = h/\Delta E$ insures that successive pulses are orthogonal to each other. With the above definitions, the calculation of the noise spectral density in the energy interval $[E - \Delta E/2, E + \Delta E/2]$ reduces to the calculation of the fluctuation in the occupation factors:

$$S(\omega = 0) = \frac{2\Delta E e^2}{\pi \hbar}(\langle g^2 \rangle - \langle g \rangle^2) \,, \tag{3.3}$$

where we have dropped the index $n$ in $g_n$ because all pulses contribute to the noise in the same fashion. Also, note that we have subtracted the average current in order to isolate the fluctuations. The calculation of the spectral density of noise is thus directly related to the statistics of the current pulses.

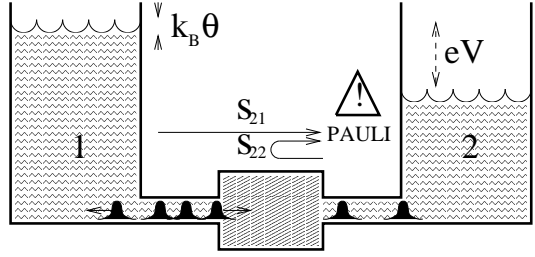

Fig. 2. The Landauer philosophy of quantum transport. Electrons travel as wave packets emitted from reservoirs (right and left). Reservoirs do not have the same chemical potential because of an applied bias. Thermal fluctuations (waves) exist in the reservoirs. The superposition of two electrons in the same scattering state cannot occur because of the Pauli principle.

To obtain the correlator $\langle g^2 \rangle - \langle g \rangle^2$, we consider all possible pulse histories: first consider the case where two electrons are incident on the sample from opposite sides. In this situation, $g = 0$ because there will be no current if the two electrons are both reflected or both transmitted, and the situation where one electron is reflected and the other is transmitted is forbidden by the Pauli principle; two electrons (with the same spin) cannot occupy the same wave packet state. Secondly, there is the straightforward situation where both incident states are empty, with $g = 0$. The other possibilities where $g = 0$ follow if an electron is reflected from one side, when no electron was incident from the other side. In fact the only possibilities where a current passes trough the sample are when an electron incident from the right (left) is transmitted while no electron was present on the other side, giving the result $g = 1$ ($g = -1$) with respective weight $f_1(1 - f_2)T$ ($f_2(1 - f_1)T$). $f_1$ ($f_2$) is the Fermi–Dirac distribution associated with the left (right) reservoir, and $T$ is the transmission probability. We therefore obtain:

$$\langle g^2 \rangle - \langle g \rangle^2 = T(f_1 + f_2 - 2f_1 f_2) - (f_1 - f_2)^2 T^2 \,. \tag{3.4}$$

A note of caution here. By removing "by hand" the two processes in which the Pauli principle is violated, it looks like the total probability for having ei-ther both electrons reflected or both electrons transmitted does not had up to one (from our argument it seems that this probability is equal to $T^2 + (1 - T)^2$. In reality, there is no such problem if one considers a wave function for incoming and outgoing states which is anti-symmetrized (as it should be). The processes where electrons are both reflected and both transmitted are not distinguishable, and their amplitudes should be added before taking the square modulus to get the total probability. Using the unitarity property of the S–matrix, one then obtain that this probability is indeed one. Summing now over all energy intervals, we thus obtain the total excess noise:

$$\begin{aligned}S(\omega = 0) &= \frac{4e^2}{h} \int dE \, T(E)[f_1(1-f_1) + f_2(1-f_2)] \\ &+ \frac{4e^2}{h} \int dE \, T(E)[1 - T(E)](f_1 - f_2)^2 \, .\end{aligned} \quad (3.5)$$

In the absence of bias or at high temperatures ($|\mu_1 - \mu_2| \ll k_B\Theta$), the two first terms on the right hand side dominate. The dependence on the distribution functions $f_i(1 - f_i)$ is typical of calculations of fluctuations in thermal equilibrium. Using the relation $f_i(1 - f_i) = -k_B\Theta\partial f_i/\partial E$, we recover the Johnson Nyquist [5] formula for thermal equilibrium noise [14]:

$$S(\omega = 0) = 4\frac{2e^2T}{h}k_B\Theta = 4Gk_B\Theta \, , \quad (3.6)$$

where $G$ is the Landauer conductance of the mesoscopic circuit. In the opposite limit, $|\mu_1 - \mu_2| \gg k_B\Theta$, we get a contribution which looks like shot noise, except that it is reduced by a factor $1 - T$:

$$S(\omega = 0) = 2e\langle I\rangle(1 - T) \, , \quad (3.7)$$

which is called reduced shot noise or quantum shot noise. At this point, it is useful to define the Fano factor as the ratio between the zero frequency shot noise divided by the Poisson noise:

$$F \equiv \frac{S(\omega = 0)}{2e\langle I\rangle} \, , \quad (3.8)$$

which is equal to $1 - T$ in the present case. In the limit of poor transmission, $T \ll 1$, and we recover the full shot noise. For highly transmissive channels, $T \sim 1$, and we can think of the deduction of shot noise as being the noise contribution associated with the poor transmission of holes across the sample. Because of the Pauli principle, a full steam of electrons which is transmitted with unit probability does not contribute to noise. Note that this is the effect seen qualitatively in point contact experiments [30, 31]. In the intermediate regime $|\mu_1 - \mu_2| \simeq k_B \Theta$, there is no clear separation between the thermal and the reduced shot noise contribution.

## 4. Generalization to multi–channel case

We now turn to the more complex situation where each lead connected to the sample has several channels. Our concern in this case is the role of channel mixing: a receiving channel on the right hand side collects electrons from all incoming channels transmitted from the left and all reflected channels on the right. We therefore expect that wave packets from these different incoming channel will interfere with each other. To avoid the issue of interference between channels and treat the system as a superposition of one dimensional contributions, we must find a wave packet representation where the mixing between channels is absent.

This representation is obtained by using a decomposition of the S– matrix describing the sample. Let us assume for simplicity that both leads have the same number of channels $M$. The S–matrix is then a block matrix containing four $M$ by $M$ sub-matrices describing the reflection from the right (left) hand side, $\mathbf{s}_{22}$ ($\mathbf{s}_{11}$), and the transmission from left to right (right to left), $\mathbf{s}_{12}$ ($\mathbf{s}_{21}$):

$$\mathbf{S} = \begin{pmatrix} \mathbf{s}_{11} & \mathbf{s}_{12} \\ \mathbf{s}_{21} & \mathbf{s}_{22} \end{pmatrix} . \tag{4.1}$$

From the unitarity of the S–matrix, which follows from current conservation, it is possible to write the sub-matrices in terms of two diagonal matrices and four unitary matrices:

$$\mathbf{s}_{11} = -i\mathbf{V}_1 \mathbf{R}^{1/2} \mathbf{U}_1^\dagger , \quad \mathbf{s}_{12} = \mathbf{V}_1 \mathbf{T}^{1/2} \mathbf{U}_2^\dagger , \tag{4.2}$$

$$\mathbf{s}_{21} = \mathbf{V}_2 \mathbf{T}^{1/2} \mathbf{U}_1^\dagger , \quad \mathbf{s}_{22} = -i\mathbf{V}_2 \mathbf{R}^{1/2} \mathbf{U}_2^\dagger , \tag{4.3}$$

where $\mathbf{R}^{1/2}$ and $\mathbf{T}^{1/2}$ are real diagonal matrices with diagonal elements $R_\alpha^{1/2}$ and $T_\alpha^{1/2}$ such that $T_\alpha + R_\alpha = 1$. $R_\alpha$ ($T_\alpha$) are the eigenvalues of the matrices $\mathbf{s}_{11}^\dagger \mathbf{s}_{11}$ ($\mathbf{s}_{21}^\dagger \mathbf{s}_{21}$). $\mathbf{U}_1, \mathbf{U}_2, \mathbf{V}_1, \mathbf{V}_2$ are unitary transformations on the eigenchannels.

Using the unitary transformations, we can now choose a new basis of incoming and outgoing states on the left and the right side of the sample: $\mathbf{U}_1$ ($\mathbf{V}_1$) is the unitary transformation used to represent the incoming (outgoing) states on the left side, while $\mathbf{U}_2$ ($\mathbf{V}_2$) is the unitary transformation used to represent the incoming (outgoing) states on the right side of the sample. The effective S–matrix thus obtained in this new basis is a block matrix of four diagonal matrices. This corresponds to a situation where no mixing between channels occurs, effectively a superposition of one-dimensional ($2 \times 2$) S–matrices which are totally decoupled (see Fig. 3). This is precisely the form which was assumed by Lesovik [17] for the adiabatic point contact. The absence of correlations between the different incoming and outgoing wave packets allows us to write the noise as a superposition of one dimensional contributions:

$$S(\omega = 0) = \frac{4e^2}{h} \int dE \sum_\alpha T_\alpha(E)[f_1(1 - f_1) + f_2(1 - f_2)]$$

$$+ \frac{4e^2}{h} \int dE \sum_\alpha T_\alpha(E)[1 - T_\alpha(E)](f_1 - f_2)^2 . \qquad (4.4)$$

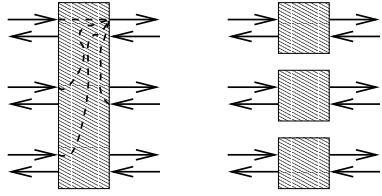

Fig. 3. A multichannel, two-terminal conductor mixes all channels in general. In the eigenchannel representation, it behaves like a set of one channel conductors which are decoupled from each other.

This expression can be cast in terms of the block elements of the initial S–matrix using the properties of the trace, $\sum_\alpha T_\alpha^n = Tr[(\mathbf{s}_{21}^\dagger \mathbf{s}_{21})^n]$. In experiments using break junctions, the $T'_\alpha s$ can actually be measured [32]: they bear the name of "mesoscopic pin code".

## 5. Scattering approach based on operator averages

The method depicted in the previous section has an intuitive character because one is able to see directly the effects of the statistics when computing the noise. It can also be generalized to treat multi-terminal conductors. In contrast, here one adopts a more systematic procedure, which is due to G. Lesovik. Soon after its appearance, this approach was generalized to multichannel, multi-terminal conductors in Ref. [19]. Here we chose to describe it because it allows to discuss noise correlations in the most straightforward way.

The philosophy of this method goes as follows. Reservoirs which are connected to the sample are macroscopic quantities. In a situation where electronic transport occurs, the chemical potential of each reservoir is only modified in a minor manner. One is thus tempted to apply equilibrium thermodynamics to quantities (operators) which involve only a given reservoir. This approach has the advantage that thermodynamical averages of operators can be computed in a systematic manner.

## 5.1. Average current

This is a classic result which can be found in Refs. [13, 33]. For simplicity, I will choose a situation where the conductor is connected to an arbitrary number of conductors, but each lead connected to the sample has only one incoming/outgoing channel. The generalization for leads containing many channels gives additional algebra.

The current operator for terminal $m$ reads in the usual way:

$$I_m(x_m) = \frac{\hbar e}{2mi} \sum_\sigma \left( \psi_{m\sigma}^\dagger(\mathbf{r}_m) \frac{\partial \psi_{m\sigma}(\mathbf{r}_m)}{\partial x_m} - \frac{\partial \psi_{m\sigma}^\dagger(\mathbf{r}_m)}{\partial x_m} \psi_{m\sigma}(\mathbf{r}_m) \right) . \tag{5.1}$$

$x_m$ is the coordinate in terminal $m$. $\psi_{m\sigma}^\dagger$ is the creation operator for a particle with spin $\sigma$ in terminal $m$. Here we shall consider conductors which transmit/reflect both spin species with the same amplitudes, and which do not scatter one spin into another along the way. In our calculations, it amounts to ignoring the spin index on the fermion operators and replacing the sum over spin by a factor 2. Spin indices will be restored in the discussion of mesoscopic superconductivity for obvious reasons. The fermion annihilation operator is expressed in terms of the scattering properties of the sample:

$$\psi_m(\mathbf{r}_m) = \sum_n \int \frac{dk_m}{\sqrt{2\pi}} \sqrt{\frac{k_m}{k_n}} (\delta_{mn} e^{ik_m x_m} + s_{mn} e^{-ik_m x_m}) c_n(k_n) . \tag{5.2}$$

$s_{mn}$ is the scattering matrix amplitude for a state incident from $n$ and transmitted into $m$. $c_m(k_m)$ is the annihilation operator for a scattering state incident from $m$. Momentum integrals are readily transformed into energy integrals with the substitution $\int dk_m \to m \int dE/\hbar^2 k_m$. Substituting the fermion operator in the expression for the current operator:

$$I_m(x_m) = \int dE \int dE' \sum_{nn'} M_m(E, E', n, n') c_n^\dagger(k_n) c_{n'}(k_{n'}) , \tag{5.3}$$

where $M_m(E, E', n, n')$ is the current matrix element, which depends on position in general. In fact, it is the sum of two distinct contributions:

$$M_m(E, E', n, n') = M_m^{\Delta k}(E, E', n, n') + M_m^{\Sigma k}(E, E', n, n') , \tag{5.4}$$

which are defined as:

$$\begin{aligned}
M_m^{\Delta k}(E, E', n, n') &= \frac{em}{2\pi\hbar^3} \left(\frac{k_m(E)k_m(E')}{k_n(E)k_{n'}(E')}\right)^{1/2} (k_m^{-1}(E) + k_m^{-1}(E')), \\
&\quad (e^{-i(k_m(E)-k_m(E'))x_m}\delta_{mn}\delta_{mn'} \\
&\quad -e^{i(k_m(E)-k_m(E'))x_m}s_{mn}^*(E)s_{mn'}(E')) \qquad (5.5) \\
M_m^{\Sigma k}(E, E', n, n') &= \frac{em}{2\pi\hbar^3} \left(\frac{k_m(E)k_m(E')}{k_n(E)k_{n'}(E')}\right)^{1/2} (k_m^{-1}(E) - k_m^{-1}(E')) \\
&\quad (-e^{-i(k_m(E)+k_m(E'))x_m}s_{mn'}(E')\delta_{mn} \\
&\quad +e^{i(k_m(E)+k_m(E'))x_m}s_{mn}^*(E)\delta_{mn'}). \qquad (5.6)
\end{aligned}$$

The current depends on the position where it is measured. This is quite unfortunate. Whether this is true of not depends on the working assumptions of our model. Here we anticipate a few results.

• When computing the average current, as our model does not take into account inelastic processes we get a delta function of energy and $M_m^{\Sigma k}(E, E', n, n') = 0$. As a result the average stationary current is constant.

• When computing the noise at zero frequency, the same thing will happen. The time integration will yield the desired delta function in energy. The zero frequency noise does not depend on where it is measured.

• When considering finite frequency noise, the term $M_m^{\Sigma k}(E, E', n, n')$ will once again be dropped out, "most of the time". Indeed, the typical frequencies which are interesting range from 0 to, say, a few $eV/\hbar$ in mesoscopic experiments. Yet in practical situations, the bias $eV/\hbar$ is assumed to be much smaller than the chemical potential $\mu_n$ of the reservoirs. Because most relevant momenta happen in the vicinity of the chemical potential within a few $eV/\hbar$, this implies that the momenta $k_m(E)$ and $k_m(E')$ are rather close. One then obtains a $M_m^{\Delta k}(E, E', n, n')$ with virtually no oscillations and a $M_m^{\Sigma k}(E, E', n, n')$ which oscillates rapidly with a wavelength $\pi/k_F$. The latter contribution has a smaller amplitude.

• About the wave vector dependences of $M_m(E, E', n, n')$. When calculating current or charge fluctuations at zero frequencies or at frequencies of the order of the bias voltage $eV/\hbar$, at the end of the day the factors containing these wave vectors cancel out in the expression of the current or noise. This is again due to the fact that $\mu_m \gg eV$.

• The question whether such oscillations are real or not depends on how the current is measured. To be reasonable, we expect that a current measurement implicitly implies an average over many Fermi wavelengths along the wire. This is the case, for instant in a Gedanken experiment where the current is detected via the magnetic field which it creates. In this case it is clear that only $M_m^{\Delta k}(E, E', n, n')$ survives.

• On the other hand, care must be taken when considering the density operator $\rho(x,t) = \Psi^\dagger(x,t)\Psi(x,t)$. This operator is related to the current operator via the continuity equation: $\partial_t \rho + \partial_x J_x = 0$ The density operator in frequency space can therefore be obtained either directly from $\Psi(x)$, or alternatively by taking derivatives of the current matrix elements. The oscillating contribution of the density operator has a matrix element of the order:

$$\rho^{\Delta k}(x_m, \omega) \sim (k_m^{-1}(E) + k_m^{-1}(E'))(k_m(E) - k_m(E')) \tag{5.7}$$

while the $2k_F$ contribution behaves like:

$$\rho^{\Sigma k}(x_m, \omega) \sim (k_m^{-1}(E) - k_m^{-1}(E'))(k_m(E) + k_m(E')) \tag{5.8}$$

As a result, they have the same order of magnitude. This is important when considering the issue of dephasing in a quantum dot close to a fluctuating current [24].

We now neglect the $2k_F$ oscillating terms and proceed. In order to compute the average current, it is necessary to consider the average:

$$\langle c_n^\dagger(k(E),t) c_{n'}(k(E'),t) \rangle = \frac{\hbar^2 k(E)}{m} f_n(E)\, \delta(E - E') \delta_{nn'} . \tag{5.9}$$

$f_n(E)$ is the Fermi-Dirac/Bose-Einstein distribution function associated with terminal $n$ whose chemical potential is $\mu_n$: $f_n(E) = 1/[\exp(\beta(E - \mu_n)) \pm 1]$. Consequently,

$$\langle I_m \rangle = \frac{2e}{h} \int dE \left( f_m(E) - \sum_n s_{mn}^*(E) s_{mn}(E) f_n(E) \right) .$$

By virtue of the unitarity of the scattering matrix, the average current becomes

$$\langle I_m \rangle = \frac{2e}{h} \sum_n \int dE\, |s_{mn}(E)|^2 \left( f_m(E) - f_n(E) \right) , \tag{5.10}$$

where $|s_{mn}(E)|^2$ is the transmission probability from reservoir $n$ to reservoir $m$. Eq. (5.10) is the Landauer formula, generalized to many channels and many terminals in Refs. [33].

*5.2. Noise and noise correlations*

The noise is defined in terms of current operators as:

$$S_{mn}(\omega) = \lim_{T \to +\infty} \frac{2}{T} \int_{-T/2}^{T/2} dt \int_{-\infty}^{+\infty} dt'\, e^{i\omega t'} \Big[ \langle I_m(t) I_n(t+t') \rangle$$

$$- \langle I_m \rangle \langle I_n \rangle \Big] . \tag{5.11}$$

This definition will be justified in the finite frequency noise section. When $m = n$, $S_{mm}$ corresponds to the (autocorrelation) noise in terminal $m$. When $m$ et $n$ differ, $S_{mn}$ corresponds to the noise cross-correlations between $m$ and $n$.

The calculation of noise involves products $I_m(t)I_n(t+t')$ of two current operators. It therefore involves grand canonical averages of four fermion operators, which can be computed with Wick's theorem:

$$\langle c^\dagger_{p_1}(k(E_1),t)c_{p_2}(k(E_2),t)c^\dagger_{p_3}(k(E_3),t+t')c_{p_4}(k(E_4),t+t')\rangle =$$
$$\frac{\hbar^4 k(E_1)k(E_3)}{m^2}f_{p_1}(E_1)f_{p_3}(E_3)\delta_{p_1 p_2}\delta_{p_3 p_4}\delta(E_1-E_2)\delta(E_3-E_4)$$
$$+\frac{\hbar^4 k(E_1)k(E_2)}{m^2}f_{p_1}(E_1)(1\mp f_{p_2}(E_3))\delta_{p_1 p_4}\delta_{p_2 p_3}$$
$$\times \delta(E_1-E_4)\delta(E_3-E_2)e^{-i(E_1-E_2)t'/\hbar} \ . \tag{5.12}$$

The first term represents the product of the average currents, while in the second term $f(1 \mp f)$, "−" corresponds to fermionic statistics, while the "+"corresponds to bosonic statistics. In the expression for the noise, only the irreducible current operator contributes, and the integral over time gives a delta function in energy (one of the energy integrals drops out). One gets the general expression for the finite frequency noise for fermions [27] :

$$S_{mn}(\omega) = \frac{4e^2}{h}\int dE \sum_{pp'}(\delta_{mp}\delta_{mp'} - s^*_{mp}(E)s_{mp'}(E-\hbar\omega))$$
$$\times (\delta_{np'}\delta_{np} - s^*_{mp'}(E-\hbar\omega)s_{mp}(E))$$
$$\times f_p(E)(1-f_{p'}(E-\hbar\omega)) \ . \tag{5.13}$$

Below, we discuss a few examples.

### 5.3. Zero frequency noise in a two terminal conductor

#### 5.3.1. General case

We can recover the result previously derived from the wave packet approach. We emphasize that these results were first derived with the present approach [17]. Consider a conductor connected to two terminals ($L$ et $R$). Considering only autocorrelation noise ($m = n$) and setting $\omega = 0$ in Eq. (5.13):

$$S_{LL}(\omega=0) = \frac{4e^2}{h}\int dE\Big[T(E)^2\left[f_L(1\mp f_L)+f_R(1\mp f_R)\right]$$
$$+ T(E)(1-T(E))\left[f_L(1\mp f_R)+f_R(1\mp f_L)\right]\Big] ,$$

$$= \frac{4e^2}{h} \sum_{\alpha=1}^{N_c} \int dE \Big[ T(E) \left[ f_L(1 \mp f_L) + f_R(1 \mp f_R) \right]$$

$$\pm T(E)(1 - T(E))(f_L - f_R)^2 \Big] . \quad (5.14)$$

The result of Eq. (3.5) is thus generalized to describe fermions or bosons. These expressions, and the corresponding limits where the voltage dominates over the temperature (shot noise) or inversely when the temperature dominates (Johnson noise) have already been discussed.

*5.3.2. Transition between the two noise regimes*

In both limits for the noise, it was assumed for simplicity that the transmission probability did not depend significantly on energy. This is justified in practice if $T(dT/dE)^{-1} \gg eV$, which can be reached by choosing both a sufficiently small voltage bias, and choosing the chemical potentials such that there are no resonances in transmission within this energy interval. In this situation, the integrals over the two finite temperature Fermi functions can be performed.

Furthermore, while the single channel results are instructive to order the fundamental features of quantum noise, in is useful to provide now the general results for conductors containing several channels. Using the eigenvalues of the transmission matrix, one then obtains [16] :

$$S_{LL}(0) = \frac{4e^2}{h} \Bigg[ 2k_B \Theta \sum_\alpha T_\alpha^2$$

$$+ eV \coth\left( \frac{eV}{2k_B \Theta} \right) \sum_\alpha T_\alpha (1 - T_\alpha) \Bigg] . \quad (5.15)$$

Of interest for tunneling situations is the case where all eigenvalues are small compared to 1. Terms proportional to $T_\alpha^2$ are the neglected: Eq. (5.15) becomes:

$$S_{LL}(0) = 2e \langle I \rangle \coth\left( \frac{eV}{2k_B \Theta} \right) . \quad (5.16)$$

*5.4. Noise reduction in various systems*

Noise reduction in a point contact was observed in semiconductor point contacts [34, 35]. For a one channel sample, the noise has a peak for transmission $1/2$, a peak which was detected in [34], and subsequent oscillations are observed as the number of channels increases. A quantitative analysis of the $1 - T$ noise reduction was performed by the Saclay group [35]. Noise reduction was subsequently observed in atomic point contacts [36, 37] using break junctions.

The $1 - T$ reduction of shot noise can be also observed in various mesoscopic systems, other than ballistic. It is most explicit in point contact experiment and break junctions experiment because one can tune the system in order to have the controlled opening of the first few conduction channels. But what happens in other systems such as cavities and multi-channel conductors ?

*5.4.1. Double barrier structures*
Double barrier structures are interesting because there exists specific energies where the transmission probability approaches unity. Inside the well, one has quasi-bound states every time that the phase accumulated in a round trip equals $2\pi$. The approximate energy dependence of the transmission coefficient for energies close to the $n^{th}$ resonance corresponds to a Breit-Wigner formula:

$$T(E) = T_n^{max} \frac{\Gamma_n^2/4}{(E - E_n)^2 + \Gamma_n^2/4} , \quad T_n^{max} = 4\frac{\Gamma_{Ln}\Gamma_{Rn}}{(\Gamma_{Ln} + \Gamma_{Rn})^2} , \quad (5.17)$$

where $\Gamma_{Ln}$ and $\Gamma_{Rn}$ are escape rates to the left and right. If the width of the individual levels are small compared to the resonance spacing, the total transmission coefficient is given by a sum of such transmission coefficients. Inserting this in the zero frequency result for the current and for the noise, one obtains current and noise contributions coming from the resonant level located in between the chemical potentials on the left and on the right. Of particular interest is the case of a single resonance:

$$\langle I \rangle = \frac{e}{\hbar} \frac{\Gamma_L \Gamma_R}{\Gamma_L + \Gamma_R} , \quad (5.18)$$

$$S = 2e \frac{(\Gamma_L^2 + \Gamma_R^2)}{(\Gamma_L + \Gamma_R)^2} \langle I \rangle = e \langle I \rangle , \quad (5.19)$$

where the last equality in Eq. (5.19) applies when the barriers are symmetric, giving a Fano factor $1/2$. Note that this approach assumes a quantum mechanical coherent treatment of transport. A remarkable fact is that the $1/2$ reduction can also be derived when transport is incoherent, using a master equation approach. [38].

*5.4.2. Noise in a diffusive conductor*
The diffusive regime can be reached with the scattering approach. This point has been emphasized previously in the context of previous les Houches summer schools [39]. Until the late 1980's, results on diffusive metals where mainly obtained using diagrammatic techniques. Ensemble averages of non-equilibrium transport properties can also be reached using the scattering approach. Random matrix theory, allows to study of both the spectrum of quantum dots and the transport properties of mesoscopic conductors. A thorough review exist on this topic [40].

Consider a conductor whose transverse dimension $W$ is larger than the Fermi wavelength: $W \gg \lambda_F$. The number of transverse channels is $N_\perp \sim W k_F/\pi \gg 1$. When disorder is included in this wire, mixing occurs between these channels: one has a mean free path $l$ which separates each elastic scattering event. When the wire is in the diffusive regime, the Fermi wave length $\lambda_F \ll l$, there are many collisions with impurities within the length of the sample $l \ll L$, while $L < L_\phi$, the phase coherence length of the sample. Due to disorder, electron states are localized within a length $L_\xi = N_\perp l$. In order to be in the metallic regime, $L \ll L_\xi$.

In quantum transport, the conductance of a diffusive metal is given by the Drude formula $\langle G \rangle \sim (n e^2 \tau/m)(W/L)$, where the precise numerical factor depends on dimensionality. At the same time, this conductance can be identified with a Landauer conductance formula which is averaged over disorder:

$$\langle G \rangle = \frac{2e^2}{h} \langle \sum_n T_n \rangle = \frac{2e^2}{h} N_\perp \langle T \rangle . \tag{5.20}$$

Identification with the Drude formula yields $\langle T \rangle = l/L \ll 1$. One could think that this implies that all channels have a low transmission: in the diffusive regime, the mean free path is much smaller than the sample length. For the noise, this would then give a Poissonian regime. On the other hand, it is known that classical conductor exhibit no shot noise. Should the same be true for a mesoscopic conductor ? The answer is that both statements are incorrect. The noise of a diffusive conductor does neither exhibit full shot noise or zero shot noise. The result is in between the two. The probability distribution of eigenvalues is bimodal:

$$P(T) = (l/2L)[T\sqrt{1-T}]^{-1} . \tag{5.21}$$

There is a fraction $l/L$ of open channels while all other channels are exponentially small (closed channels). Consider now the noise, averaged over impurities:

$$\langle S(\omega = 0) \rangle = \frac{4e^2}{h} N_\perp \langle T(1-T) \rangle . \tag{5.22}$$

The random matrix theory average yields [41] $\langle T(1-T) \rangle = l/3L$, so that the Fano factor is $F = 1/3$. This result has an universal character: it does not depend of the sample characteristics. It should be also noted that it can be recovered using alternative approaches, such as the Boltzmann-Langevin semi-classical approach [42,43]. This shot noise reduction was probed experimentally in small metallic wires [44]. A careful tuning of the experimental parameters (changing the wire length and the geometry of the contacts) allowed to make a distinction between this disordered induced shot noise reduction ($1/3$), and the $\sqrt{3}/4$ reduction associated with hot electrons.

*5.4.3. Noise reduction in chaotic cavities*

Chaotic cavities were studied by two groups [45, 46], under the assumption that the scattering matrix belongs to the so called Dyson circular ensemble. This is a class of random matrix which differs from the diffusive case, and the probability distribution for transmission eigenvalues is given by:

$$P(T) = 1/[\pi\sqrt{T(1-T)}]. \tag{5.23}$$

This universal distribution has the property that $\langle T \rangle = 1/2$ and $\langle T(1-T) \rangle = 1/8$, so that the Fano factor is $1/4$. This result applies to open cavities, connected symmetrically (same number of channels) to a source and drain. It is nevertheless possible [48] to treat cavities with arbitrary connections to the reservoirs, which allows to recover the $1/2$ suppression in the limit of cavities with tunneling barriers. The shot noise reduction in chaotic cavities was observed in gated two dimensional electron gases [47], where it was possible to analyze the effect of the asymmetry of the device (number of incoming/outgoing channels connected to the cavity).

## 6. Noise correlations at zero frequency

*6.1. General considerations*

It is interesting to look at how the current in one lead can be correlated to the current in another lead. Let us focus on zero frequency noise correlations for fermions, at zero temperature. In this case, the Fermi factors enforce in Eq. (5.13) that the contribution from $p = p'$ vanishes. The remaining contribution has matrix elements $M_m(E, p, p') \sim s^*_{mp} s_{mp'}$, because the delta function term drops out. The correlation then reads:

$$S_{mn}(\omega = 0) = \frac{2e^2}{h} \sum_{pp'} \int dE\, s^*_{mp} s_{mp'} s^*_{np'} s_{np} f_p(E)(1 - f_{p'}(E)). \tag{6.1}$$

The contribution $p = p'$ has been kept in the sums, although it vanishes, for later convenience. From this expression it is clear that the noise autocorrelation $m = n$ is always positive, simply because scattering amplitudes occur next to their hermitian conjugates.

On the other hand, if $m \neq n$, the correlations are negative. The unitarity of the S matrix implies in this case $\sum_p s_{mp} s_{np}^* = 0$, so that the terms linear in the Fermi functions drop out:

$$S_{mn}(\omega = 0) = -\frac{2e^2}{h} \int dE \left| \sum_p s_{np} s_{mp}^* f_p(E) \right|^2 . \tag{6.2}$$

If one wishes to measure the cross correlations between $m$ and $n$, one could imagine fabricating a new lead "$m+n$" (assuming that both $m$ and $n$ are at the same chemical potential), and measuring the autocorrelation noise $S_{(m+n)(m+n)}$. The cross correlation is then, obtained by subtracting the autocorrelation of each sub-lead, and dividing by two:

$$S_{(m+n)}(\omega = 0) = S_m(\omega = 0) + S_n(\omega = 0) + 2 S_{mn}(\omega = 0) . \tag{6.3}$$

This prescription was used in the wave packet approach to analyze the Hanbury-Brown and Twiss correlations for fermions [16], which are discussed below.

### 6.2. Noise correlations in a Y–shaped structure

In 1953, Hanbury–Brown and Twiss performed several experiments [25] where two detectors at different locations collected photons emitted from a light source. Their initial motivation was to measure the size of distant star. This experiment was followed by another one where the source was replaced by a mercury arc lamp with filters. The filters insured that the light was essentially monochromatic, but the source was thermal. A semi-transparent mirror split the beam in two components, which were fed to two photo-multiplier tubes. The correlations between the two detectors where measured as a function of the distance separating the detectors, and where found to be always positive. The experiment was subsequently explained to be a consequence of the bunching effect of photons. Because the light source was thermal, several photons on average occupied the same transverse states of the beams, so the detection of one or several photons in one arm of the beam was typically accompanied by the detection of photons in the other arm.

This measurement, which can be considered one of the first in the field of quantum optics, can thus be viewed as a check that photons are indistinguishable particles which obey Bose–Einstein statistics. It turns out that this result is also fully consistent with a classical electromagnetism description: the photon bunching effect can be explained as a consequence of the superposition principle for light applied to noisy sources, because the superposition principle follows from Bose Einstein statistics.

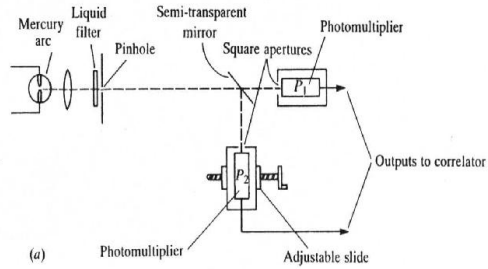

Fig. 4. The Hanbury-Brown and Twiss experiment.

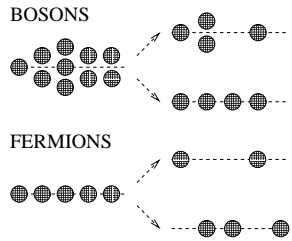

Fig. 5. Illustration of the bunching effect for fermions, and of the anti-bunching effect for electrons in a Hanbury-Brown and Twiss geometry.

Nowadays, low flux light sources – which are used for quantum communication purposes – can be achieved such that photons are emitted one by one. The resulting correlation signal is then negative because the detection of a photon in one arm means that no photon is present is the other arm. This is in fact what happens for electrons.

It has been suggested by many authors that the analog experiment for fermions should be performed [49]. In this way the Fermi Dirac statistics of electrons could be diagnosed directly from measuring the noise correlations of fermions passing through a beam splitter. Because of the Pauli principle, two electrons cannot occupy the same transverse state. The correlations should then be be negative. For technical reasons – the difficulty for achieving a dense beam of electrons in vacuum – the experiment proved quite difficult to achieve. However, it was suggested [16] that a similar experiment be performed for fermions in nanostructures.

Consider a three terminal conductor, a "Y junction". Electrons are injected from terminal 3, which has a higher chemical potential than terminals 1 and 2, where the correlations are measured. Here we focus on zero frequency noise (setting $\omega = 0$ in Eq. (5.13)). For simplicity we choose to work at zero temperature. The autocorrelation noise becomes:

$$S_{mm} = \frac{4e^2}{h} \sum_{p \neq p'} |s_{mp}|^2 |s_{mp'}|^2 f_p(E)(1 - f_{p'}(E)) .$$

Assuming that $\mu_1 = \mu_2$, the voltage bias is then $eV = \mu_3 - \mu_{1,2}$. and the autocorrelation becomes $S_{11} = (4e^2/h)eVT_{13}(1 - T_{13})$, and similarly for terminal 2. On the other hand, the correlations between 1 and 2 yield:

$$S_{12} = = -\frac{4e^2}{h} eV |s_{13}|^2 |s_{23}|^2 . \tag{6.4}$$

A natural way to summarize these results is to normalize the correlations by the square root of the product of the two autocorrelations:

$$S_{12}/\sqrt{S_{11} S_{22}} = -\sqrt{T_{13} T_{23}}/\sqrt{(1 - T_{13})(1 - T_{23})} , \tag{6.5}$$

where $T_{13}$ and $T_{23}$ are the transmission probabilities from 3 to 1 and 2. The correlations are therefore negative regardless of the transmission of the sample. Quantitative agreement with experiments has been found in the late nineties in two separate experiments. A first experiment [50] designed the electron analog of a beam splitter using a thin metallic gate on a two dimensional electron gas. A second experiment was carried out in the quantum Hall effect regime [51]: a point contact located in the middle of the Hall bar then plays the role of a controllable beam splitter, as the incoming edge state is split in two outgoing edge states at its location. Recently, negative noise correlations have been observed in electron field emission experiments [52].

## 7. Finite frequency noise

### 7.1. Which correlator is measured ?

Finite frequency noise is the subject of a debate. What is actually measured in a finite frequency noise measurements ? The current operator is indeed an hermitian operator, but the product of two current operators evaluated at different times is not hermitian. If one takes the wisdom from classical text books [53], one is told that when one is faced with the product of two hermitian operators, one should symmetrize the result in order to get a real, measurable quantity. This procedure has led to a formal definition of finite frequency noise [27]:

$$S_{sym}(\omega) = \int dt \, e^{i\omega t} \langle\langle I(t)I(0) + I(0)I(t)\rangle\rangle . \tag{7.1}$$

(the double bracket means the product of averages current has been subtracted out). At the same time, one can define two unsymmetrized correlators:

$$S_+(\omega) = 2\int dt\, e^{i\omega t}\langle\langle I(0)I(t)\rangle\rangle\,. \tag{7.2}$$

$$S_-(\omega) = 2\int dt\, e^{i\omega t}\langle\langle I(t)I(0)\rangle\rangle\,. \tag{7.3}$$

The factor two has been added in order to be consistent with the Schottky relation. Note that in these expressions, the time dependence is specified by the Heisenberg picture. Assuming that one knows the initial (ground) states $|i\rangle$ and the final states $|f\rangle$, one concludes that:

$$S_+(\omega) = 4\pi \sum_{if} |\langle f|I(0)|i\rangle|^2 P(i)\delta(\omega + E_f - E_i)\,. \tag{7.4}$$

$$S_-(\omega) = 4\pi \sum_{if} |\langle f|I(0)|i\rangle|^2 P(i)\delta(\omega + E_i - E_f)\,. \tag{7.5}$$

Therefore, in $S_+(\omega)$ (in $S_-(\omega)$), positive (negative) frequencies correspond to an emission rate from the mesoscopic device, while negative (positive) frequencies correspond to an absorption rate. Since one does not expect to be able to extract energy from the ground state of this device, one concludes that the physically relevant frequencies for $S_+(\omega)$ (for $S_-(\omega)$) are $\omega > 0$ ($\omega < 0$).

*7.2. Noise measurement scenarios*

Ref. [54] argues that one has to specify a measurement procedure in order to decide which noise correlator is measured. In their proposal (Fig. 6a), the noise is measured by coupling the mesoscopic circuit (the antenna) inductively to a LC circuit (the detector). By measuring the fluctuations of the charge on the capacitor, one obtains a measurement of the current-current fluctuations which is is weighted by Bose-Einstein factors, evaluated at the resonant frequency $\omega = 1/\sqrt{LC}$ of the circuit:

$$\langle\langle Q^2(0)\rangle\rangle = K[S_+(\omega)(N_\omega + 1) - S_-(\omega)N_\omega]\,, \tag{7.6}$$

where $K$ is a constant which depends on the way the two circuits (antenna and detector) are coupled and the double brackets imply that one measures the excess charge fluctuations. $N_\omega$ are bosonic occupation factors for the quantized LC circuit. It is therefore a mixture of the two unsymmetrized noise correlators which is measured in this case. This point has been reemphasized recently [55].

Another proposal considers [56] a capacitive coupling between the mesoscopic circuit and the detector circuit (Fig. 6b). The detector circuit consists

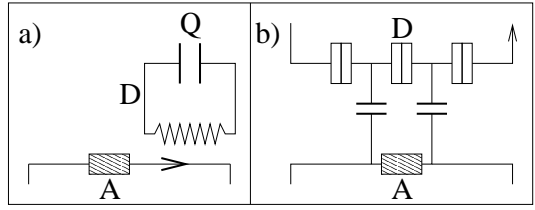

Fig. 6. Schematic description of noise measurement setups. $A$ (antenna) is the mesoscopic circuit to be measured, while $D$ is the detector circuit. a) inductive coupling with an $LC$ circuit. b) capacitive coupling with a double dot

of a double dot system embedded in a a circuit where current is measured. Each dot is coupled to each side of the mesoscopic device: voltage fluctuations in this antenna circuit are translated into voltage fluctuations – thus phase fluctuations – between the two dots. Indeed, it is understood since the early nineties that in nano-junctions, that the phase is the canonical conjugate of the charge in the junction. The Fermi golden rule calculation of the current across the junction needs to be revisited in order to take into account the effect of the environment. This is the so-called $P(E)$ theory of the dynamical Coulomb blockade [57]. $P(E)$ is the probability distribution for inelastic scattering: it is equal to a delta function for a low impedance environment. In the general case it is specified as follows:

$$P(\epsilon) = h^{-1} \int_{-\infty}^{+\infty} \exp\left[J(t) + i\frac{\epsilon}{\hbar}t\right] dt \;. \tag{7.7}$$

Here, $J(t)$ is the (unsymmetrized) correlation function of the phase. In Aguado and Kouvenhoven's proposal, the impedance environment coupled to the double dot does not only consist of the leads connected to these (these leads are assumed to be massive and well conducting): the environment is the mesoscopic circuit itself. Letting $\epsilon$ denote the level separation between the two dots, a DC inelastic current can circulate in the detection circuit only if the frequency $\omega = \epsilon/\hbar$ is provided by the mesoscopic circuit (antenna). The inelastic current is then given by

$$I_D(\epsilon) = \frac{e}{\hbar} T_c^2 P(\epsilon) \;, \tag{7.8}$$

where $T_c$ is the tunnel amplitude between the dots. In $P(\epsilon)$, the phase correlator $J(t)$ contains the trans-impedance $Z_{trans}(\omega)$ connecting the detector and the antenna circuit, as well as the unsymmetrized noise:

$$J(t) = \frac{e^2}{2\hbar^2} \int_{-\infty}^{+\infty} \frac{|Z_{trans}(\omega)|^2}{\omega^2} S_-(\omega)(e^{-i\omega t} - 1) \;. \tag{7.9}$$

Under the assumption of a low impedance coupling, the inelastic current becomes:

$$I_D(\epsilon) \simeq 2\pi^2 \kappa^2 \frac{T_c^2}{e} \frac{S_-(\epsilon/\hbar)}{\epsilon} \ . \tag{7.10}$$

Here, $\kappa$ depends on the parameters (capacitances, resistors,...) of both circuits. Given that the energy spacing between the dots can be controlled by a gate voltage, the measurement of $I_D$ provides a direct measurement of the noise – non-symmetrized – of the mesoscopic device.

This general philosophy has been tested recently by the same group, who measured [58] the high frequency noise of a Josephson junction using another device, a – superconductor-insulator-superconductor junction – as a detector: in this system, quasiparticle tunneling in the SIS junction can occur only if it is assisted by the frequency provided by the antenna.

### 7.3. Finite frequency noise in point contacts

Concerning the finite frequency noise, results are better illustrated by choosing the zero temperature case, and keeping once again the assumption that the transmission amplitudes are constant. In this case, only the last term in Eq. (5.13) contributes, and one obtains the result [17, 59]:

$$S_+(\omega) = \frac{4e^2}{h} T(1-T)(eV - \hbar\omega) \ \Theta(eV - \hbar\omega) \ . \tag{7.11}$$

The noise decreases linearly with frequency until $eV = \hbar\omega$, and vanishes beyond that. The noise therefore contains a singularity: its derivative diverges at this point. Properly speaking, the above result is different from that of the first calculation of finite frequency noise [59], where the noise expression was symmetrized with respect to the two time arguments. It coincides only at the $\omega = 0$ point, and in the location of the singularity. This singularity was first detected in a photo-assisted transport measurement, using a metallic wire in the diffusive regime [60]. Photo-assisted shot noise is briefly discussed in the conclusion as an alternative to finite frequency noise measurements

In what follows, when talking about finite frequency noise in superconducting–normal metal junctions, the unsymmetrized correlator $S_+(\omega)$ will always be considered.

## 8. Noise in normal metal-superconducting junctions

Before mesoscopic physics was born, superconductors already displayed a variety of phase coherent phenomenon, such as the Josephson effect. Scientists be-gan wondering what would happen if a piece of coherent, normal metal, was put in contact with a superconductor. The phenomenon of Andreev reflection [61] – where an electron incident from a normal metal on a boundary with a superconductor is reflected as a hole (Fig. 7), now has its share of importance in mesoscopic physics. In electron language, Andreev reflection also corresponds to the absorption of a two electrons as a Cooper pair in the superconductor. Here we analyze the noise of normal metal–superconductor junctions. Such junctions bear strong similarities with two terminal, normal conductors, except that there are two types of carriers. While charge is not conserved in an Andreev process, energy is conserved because the two electrons incident on the superconductor – one above the Fermi level and the other one symmetrically below – have the total energy for the creation of a Cooper pair, which is the superconductor chemical potential. An NS junction can be viewed as the electronic analog of a phase conjugation mirror in optics. Finally, spin is also conserved at the boundary: the electron and its reflected hole have opposite spin, because in the electron picture, the two electrons entering the superconductor do so as a singlet pair because of the s wave symmetry of the superconductor. This fact will be important later on when discussing entanglement.

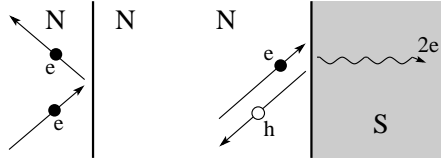

Fig. 7. Left, normal reflection: an electron is reflected as an electron, and parallel momentum is conserved. Right, Andreev reflection: an electron is reflected as a hole whose momentum is exactly opposite of that of the electron. A charge $2e$ is absorbed in the superconductor as a Cooper pair.

### 8.1. Bogolubov transformation and Andreev current

Consider a situation where a superconductor is connected to several normal metal leads, in turn connected to reservoirs with chemical potentials $\mu_m$. For simplicity, we assume each lead to carry only one single transverse channel. All energies in the leads are measured with respect to the superconductor chemical potential $\mu_S$. If one studies transport from the point of view of scattering theory, the Bogolubov de Gennes theory of inhomogeneous superconductors [62, 63] is best suited.

The starting point is the mean field Hamiltonian of Bogolubov which describes a system of fermions subject to a scalar potential and an attractive interaction. This latter interaction contains in principle two creation operators as well as two annihilation operators, rendering it a difficult problem to solve. Bogolubov had the originality to propose an effective Hamiltonian which does not conserve particles [63]. This Hamiltonian is diagonalized by the Bogolubov transformation:

$$\psi_{i\sigma}(x) = \sum_{j\beta} \int \frac{dk}{\sqrt{2\pi}} \left[ u_{ij\beta}(x)\gamma_{j\beta\uparrow}(k) - \sigma v^*_{ij\beta}(x)\gamma^+_{j\beta\downarrow}(k) \right] . \quad (8.1)$$

The state $u_{ij\beta}$ ($v_{ij\beta}$) corresponds to the wave function of a electron-like (hole-like) quasiparticle in terminal $i$ injected from terminal $j$ as a quasiparticle $\beta$ ($\beta = e, h$). $\gamma(k)$ et $\gamma^\dagger(k)$ are fermionic quasiparticle operators. As before it will be convenient to switch to energy integrals with the substitution $\int dk = \int dE/\sqrt{\hbar v^j_{e,h}(E)}$, i.e., electrons and holes do not have the same velocities. The corresponding Hamiltonian has a diagonal form:

$$H_{eff} = \sum_{j\beta\sigma} \int_0^{+\infty} dE\, E\, \gamma^+_{j\beta\sigma}(E)\gamma_{j\beta\sigma}(E) , \quad (8.2)$$

provided that the electron and hole wave functions satisfy:

$$\begin{cases} Eu_{ij\beta} = \left( -\frac{\hbar^2}{2m}\frac{\partial^2}{\partial x^2} - \mu_S + V(x) \right) u_{ij\beta} + \Delta(x)v_{ij\beta} , \\ \\ Ev_{ij\beta} = -\left( -\frac{\hbar^2}{2m}\frac{\partial^2}{\partial x^2} - \mu_S + V(x) \right) v_{ij\beta} + \Delta^*(x)u_{ij\beta} . \end{cases} \quad (8.3)$$

In principle, these equations have to be solved self-consistently as the gap parameter depends on $u_{ij\beta}$ and $v_{ij\beta}$. In most applications however, the gap is assumed to be a step-wise function describing an abrupt transition from a superconductor to a normal metal lead. On the normal metal side, the Bogolubov-de Gennes equations (8.3) can be solved easily assuming plane wave solutions for normal metals and holes. For a given energy $E$, the corresponding wave numbers are $k_e^N = \sqrt{2m(\mu_S + E)}/\hbar$ and $k_h^N = \sqrt{2m(\mu_S - E)}/\hbar$. We are now dealing with a S-matrix which can either convert electrons from terminal $j$ to terminal $i$, or electrons into holes in these terminals:

$$u_{ij\beta}(x) = \delta_{i,j}\delta_{e,\beta}e^{ik_e^N x} + s_{ije\beta}\sqrt{\frac{k_\beta^j}{k_e^N}}e^{-ik_e^N x} , \quad (8.4)$$

$$v_{ij\beta}(x) = \delta_{i,j}\delta_{h,\beta}e^{-ik_h^N x} + s_{ijh\beta}\sqrt{\frac{k_\beta^j}{k_h^N}}e^{ik_h^N x} . \quad (8.5)$$

A particular aspect of this formalism is that electrons and holes have opposite momenta. $s_{ij\alpha\beta}$ is the amplitude for getting a particle $\alpha$ in terminal $i$ given that a particle $\beta$ was incident from $j$. In comparison to the previous section, the spin index has to be restored in the definition of the current operator:

$$
\begin{aligned}
I_i(x) &= \frac{e\hbar}{2miv_F} \frac{1}{2\pi\hbar} \int_0^{+\infty} dE_1 \int_0^{+\infty} dE_2 \sum_{mn\sigma} \\
&\quad \Bigl[ \Bigl( u_{im}^* \partial_x u_{in} - (\partial_x u_{im}^*) u_{in} \Bigr) \gamma_{m\sigma}^\dagger \gamma_{n\sigma} \\
&\quad - \Bigl( u_{im}^* \partial_x v_{in}^* - (\partial_x u_{im}^*) v_{in}^* \Bigr) \sigma \, \gamma_{m\sigma}^\dagger \gamma_{n-\sigma}^\dagger \\
&\quad - \Bigl( v_{im} \partial_x u_{in} - (\partial_x v_{im}) u_{in} \Bigr) \sigma \, \gamma_{m-\sigma} \gamma_{n\sigma} \\
&\quad + \Bigl( v_{im} \partial_x v_{in}^* - (\partial_x v_{im}) v_{in}^* \Bigr) \gamma_{m-\sigma} \gamma_{n-\sigma}^\dagger \Bigr] \,.
\end{aligned} \quad (8.6)
$$

In order to avoid the proliferation of indices, we chose to replace sums over $j$ (terminal number) and $\beta$ (electron or hole) by a single index $m$. Expressions containing $m$ $n$ also have an energy dependence $E_1$ ($E_2$). In the following calculations, the energy dependence of the wave numbers is neglected for simplicity. As before, this is justified from the fact that all chemical potentials (normal leads and superconductors) are large compared to the applied biases. Note that unlike the normal metal case, the current operator does not conserve quasiparticles. The average current implies the computation of the average $\langle \gamma_{m\sigma}^\dagger(E_1) \gamma_{n\sigma}(E_2) \rangle = f_{j\alpha}(E_1) \delta_{mn} \delta(E_1 - E_2)$. The distribution function $f_m$ depends on which type of particle is considered: it is the Fermi Dirac distribution for electron in a normal lead $f_{m(e)}(E) = 1/[1 + e^{\beta(E-\mu_j)}]$; for holes in the same lead it represents the probability for a state with energy $-E$ to be empty $f_{m(h)}(E) = 1 - f_{m(e)}(-E) = 1/[1 + e^{\beta(E+\mu_m)}]$; for electron or hole-like quasiparticles in the superconductor, it is simply $f_S(E) = 1/[1 + e^{\beta E}]$. The average current in lead $i$ becomes:

$$
\langle I_i(x) \rangle = \frac{e}{2\pi m i v_F} \int_0^{+\infty} dE \sum_m \Bigl[ \Bigl( u_{im}^* \partial_x u_{im} - (\partial_x u_{im}^*) u_{im} \Bigr) f_m \\
+ \Bigl( v_{im} \partial_x v_{im}^* - (\partial_x v_{im}) v_{im}^* \Bigr) (1 - f_m) \Bigr] \,. \quad (8.7)
$$

For a single channel normal conductor connected to a superconductor, so the Andreev regime means that the applied bias is much smaller than the superconducting gap, so that no quasiparticles can be excited in the transport process. It is also assumed that the scattering amplitudes have a weak energy dependence within the range of energies specified by the bias voltage. Using the unitarity of the S–matrix one obtains:

$$\langle I \rangle = \frac{4e^2}{h} R_A V , \qquad (8.8)$$

where $R_A = |s_{11he}|^2$ is the Andreev reflection probability. The conductance of a normal metal-superconductor junction in then doubled [64, 65] because of the transfer of two electron charges. Indeed, this result could have been guessed from a simple extension of the Landauer formula to NS situations.

*8.2. Noise in normal metal–superconductor junctions*

For the Andreev regime, in a single NS junction, noise can then be calculated using the wave packet approach [66], with the following substitutions from the normal metal case, Eq. (5.14):
• the transmission probability is replaced by the Andreev reflection probability: $T \to R_A$.
• The transfered charge is $2e$.
• Electrons have a Fermi distribution $f(E - eV)$, while holes have a Fermi distribution $f(E + eV)$.
• Although electrons with spin $\sigma$ are converted into holes with spin $-\sigma$, the spin index is ignored, which is justified if the normal lead is non-magnetic. Here the spin only provides a factor two.
Using the general "wave packet" formula for the two terminal noise of normal conductors, one readily obtains:

$$\begin{aligned} S(0) &= \frac{8e^2}{h} \int dE \, \big[ R_A(E)(1 - R_A(E))(f(E - eV) - f(E + eV))^2 \\ &\quad + R_A(E)[f(E - eV)(1 - f(E - eV)) \\ &\quad + f(E + eV)(1 - f(E + eV))]\big] , \end{aligned} \qquad (8.9)$$

which (with the same standard assumptions) yields immediately the two known limits. For a voltage dominated junction $eV \gg k_B \Theta$,

$$S(0) = \frac{16e^3}{h} R_A (1 - R_A) V \equiv 4e \langle I \rangle (1 - R_A) , \qquad (8.10)$$

while in a temperature dominated regime

$$S(0) = \frac{16e^2}{h} R_A k_B \Theta \equiv 4 G_{NS} k_B \Theta , \qquad (8.11)$$

with $G_{NS}$ the conductance of the NS junction, and one recovers the fluctuation dissipation theorem.

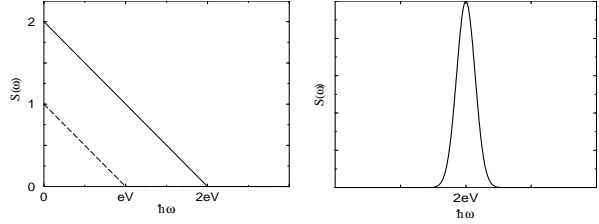

Fig. 8. Noise as a function of frequency. Left, full line: the noise (in units of $e\langle I\rangle(1-R_A)$)in an NS junction has a singularity at $\omega = 2eV/\hbar$. Left, dashed line: noise (in units of $e\langle I\rangle(1-T)$) for a junction between two normal metals with a singularity at $\hbar\omega = eV$. Right: noise in a Josephson junction, which presents a peak at the Josephson frequency. The line-width is due to radiation effects.

One now needs a general description which can treat finite frequencies, and above gap processes, (when for instance an electron is transfered into the superconductor as an electron quasiparticle). The scattering formalism based on operator averages is thus used.

In order to guess the different contributions for the noise correlator, consider the expression for $I_i(x,t)I_j(x,t+t')$. It is a product of four quasiparticle creation/annihilation operators. It will have non-zero average only if two $\gamma$ are paired with two $\gamma^\dagger$. The same is true for the noise in normal metal conductors. Here, however, electron-like and hole-like contributions will occur, but the current operator of Eq. (8.6) also contains terms proportional to $\gamma\gamma$ and $\gamma^\dagger\gamma^\dagger$ which contribute to the noise. To compute the average $\langle I_i(x,t)I_j(x,t+t')\rangle$ is useful to introduce the following matrix elements:

$$\begin{aligned}
A_{imjn}(E,E',t) &= u_{jn}(E',t)\partial_x u^*_{im}(E,t) - u^*_{im}(E,t)\partial_x u_{jn}(E',t)\,, \\
B_{imjn}(E,E',t) &= v^*_{jn}(E',t)\partial_x v_{im}(E,t) - v_{im}(E,t)\partial_x v^*_{jn}(E',t)\,, \\
C_{imjn}(E,E',t) &= u_{jn}(E',t)\partial_x v_{im}(E,t) - v_{im}(E,t)\partial_x u_{jn}(E',t)\,.
\end{aligned}$$

The two first matrix elements involve products of either particle or hole wave function. Compared to the normal case, the last one, $C_{imjn}(E,E',t)$ is novel, because it involves a mixture of electrons and holes. It will be important in the derivation of the finite frequency spectrum of noise correlations. Computing the grand canonical averages, one obtains the difference $\langle I_i(t)I_j(t+t')\rangle - \langle I_i\rangle\langle I_j\rangle$, and the Fourier transform is performed in order to compute the noise. The integration over $t'$ gives a delta function in energy as before. Note that because we are assuming positive frequencies, terms proportional to $(1-f_m)(1-f_n)$ vanish.

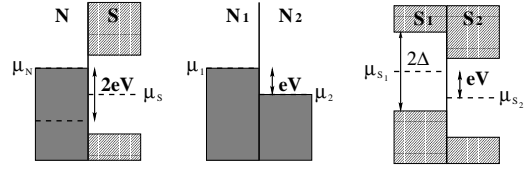

Fig. 9. Energy diagrams for three different types of junctions: left, normal metal ($\mu_N$) connected to a superconductor ($\mu_S$) ; center, 2 normal metals with chemical potentials $\mu_1$ et $\mu_2$ ; right, 2 superconductors (Josephson junction), with chemical potentials $\mu_{S_1}$ and $\mu_{S_2}$.

The noise cross-correlations have the general form:

$$
\begin{aligned}
S_{ij}(\omega) &= \frac{e^2 \hbar^2}{m^2 v_F^2} \frac{1}{2\pi\hbar} \int_0^{+\infty} dE \sum_{m,n} \\
&\quad \Bigg\{ \Theta(E+\hbar\omega) f_m(E+\hbar\omega)(1-f_n(E)) \\
&\quad \times \big| A_{imjn}(E+\hbar\omega, E) + B^*_{imjn}(E+\hbar\omega, E) \big|^2 \\
&\quad + \Theta(\hbar\omega - E) f_m(\hbar\omega - E) f_n(E) C^*_{imjn}(\hbar\omega - E, E, t) \\
&\quad \times \Big[ C_{jnim}(E, \hbar\omega - E) + C_{imjn}(\hbar\omega - E, E) \Big] \Bigg\}.
\end{aligned}
$$

Terms proportional to $f_n f_m$ $(1-f_n)(1-f_m)$ disappear, unless the frequency is non-zero, because of the energy requirements. The zero frequency limit can thus be written in a concise form [27, 67, 68]:

$$
\begin{aligned}
S_{ij}(0) &= \frac{e^2 \hbar^2}{m^2 v_F^2} \frac{1}{2\pi\hbar} \int_0^{+\infty} dE \sum_{m,n} f_m(E)(1-f_n(E)) \\
&\quad \times \big| A_{imjn}(E, E) + B^*_{imjn}(E, E) \big|^2 .
\end{aligned} \tag{8.12}
$$

### 8.3. Noise in a single NS junction

From Eq. (8.12), choosing zero temperature and using the expressions of $A_{mn}$, $B_{mn}$ and $C_{mn}$, once has to consider separately the three frequency intervals $\hbar\omega < eV$, $eV < \hbar\omega < 2eV$ and $\hbar\omega > 2eV$. The first two intervals give a noise contribution, while the last one yields $S(\omega) = 0$. Particularly puzzling is the fact that one needs to separate two regimes in frequency, while it is expected that the frequency $eV = \hbar\omega$ should no show any particular features in the noise.

*8.3.1. Below gap regime*

First consider the case where $eV \ll \Delta$. We also assume the scattering amplitudes to be independent on energy (this turns out to be a justified assumption in BTK model which we discuss shortly). One then obtains [69, 70]:

$$S(\omega) = \frac{8e^2}{\hbar}(2eV - \hbar\omega)R_A(1 - R_A)\Theta(2eV - \hbar\omega) . \tag{8.13}$$

Just as in the normal case, the noise decreases linearly with frequency, and vanishes beyond the Josephson frequency $\omega_J = 2eV/\hbar$ (left of Fig. 8). There is thus a singularity in the noise at that particular frequency. This result should be compared first to the normal case [17, 59], yet it should also be compared to the pioneering work on the Josephson junction [71, 72].

In the pure normal case (Fig. 9, center), wave functions have a time dependence $\psi_{1,2} \sim \exp[-i\mu_{1,2}t/\hbar]$, so while the resulting current is stationary, finite frequency noise contains the product $\psi_1\psi_2^*$ which gives rise to the singularity at $|\mu_2 - \mu_1|/\hbar = eV/\hbar$.

In the pure superconductor case (Fig. 9, right) an constant applied bias generates an oscillating current. $\psi_{1,2} \sim \exp[-i2\mu_{S_{1,2}}t/\hbar]$, so the order parameter oscillates as where $\mu_{S_1}$ and $\mu_{S_2}$ are the chemical potentials of the two superconductors. The noise characteristic has a peak at the Josephson frequency whose line-width was first computed in the sixties [72].

Turning back to the case of an hybrid NS junction, because here the bias is smaller than the gap, the Andreev process is the only process available to transfer charge, which allows Cooper pairs to be transmitted in or emitted from the superconductor. An electron incident from the normal side with energy $\mu_S + eV$ gets paired with another electron with energy $\mu_S - eV$, thus this second electron corresponds to a hole at $\mu_S - eV$. Both electrons have a total energy $2\mu_S$, which corresponds to the formation of a Cooper pair. The incoming electron wave function oscillates as $\psi_e \sim \exp[-i(\mu_S + eV)t/\hbar]$, while the hole wave function oscillates as $\psi_h \sim \exp[-i(\mu_S - eV)t/\hbar]$ (figure 9 left). The noise therefore involves the product $\psi_e\psi_h^*$ which oscillates at the Josephson frequency, thus giving rise to the singularity in the noise derivative. This can be considered as an analog of the Josephson effect observed in a single superconductor adjacent to a normal metal, but only in the noise.

*8.3.2. Diffusive NS junctions*

We have seen that the zero frequency shot noise of a tunnel junction is doubled [28, 66, 73, 74]. It is also interesting to consider a junction between a diffusive normal metal on one side, in perfect contact with a superconductor. The junction contains many channels, yet one can also find an eigenchannel representation where the noise is expressed as a linear superposition of independent single-channel junctions. Consider here current and shot noise at zero temperature:

$$\langle I \rangle = \frac{4e^2}{h} \sum_n R_{A_n}, \qquad (8.14)$$

$$S(\omega = 0) = \frac{16e^2}{h} \sum_n R_{A_n}(1 - R_{A_n}). \qquad (8.15)$$

In order to compute the quantity $\langle R_{A_n}(1 - R_{A_n}) \rangle$ using random matrix theory, a specific model for the NS junction has to be chosen. A natural choice [40] consists of a normal disordered region separated from a perfect Andreev interface. Because the ideal Andreev interface does not mix the eigenchannels, the Andreev reflection eigenvalues can be expressed in terms of the transmission eigenvalues of the normal metal scattering matrix which models the disordered region: $R_{A_n} = T_n^2/(2 - T_n)^2$. The noise can in turn be expressed in terms of these eigenvalues:

$$S(\omega = 0) = \frac{64e^2}{h} \sum_n \frac{T_n^2(1 - T_n)}{(2 - T_n)^4}. \qquad (8.16)$$

Note that channels with either high or low transmission do not contribute to the shot noise. First, assume that all channels have the same transmission probability $\Gamma$. $\Gamma$ represents the transparency per mode of the NS interface, but no mixing is assumed between the modes. The noise can be written in this case as:

$$S = \frac{8(1 - \Gamma)}{(2 - \Gamma)^2}(2e\langle I \rangle). \qquad (8.17)$$

For $\Gamma \ll 1$, one obtains a Poissonian noise of uncorrelated charges $2e$. This means that the shot noise is doubled compared to the normal tunnel barrier shot noise result. Next, one considers a disordered normal region with an ideal interface. The averages over transmission eigenvalues are computed using random matrix theory:

$$S(\omega = 0) = \frac{2}{3}(2e\langle I \rangle). \qquad (8.18)$$

Thus giving a 2/3 reduction for the disordered NS interface. The above doubling of the shot noise was observed in S-N-S junctions [75], and in single NS junctions [76,77]. The data on the detection of the $2e$ charge was obtained by the Grenoble and by the Yale group, the latter in the measurement of photo-assisted shot noise.

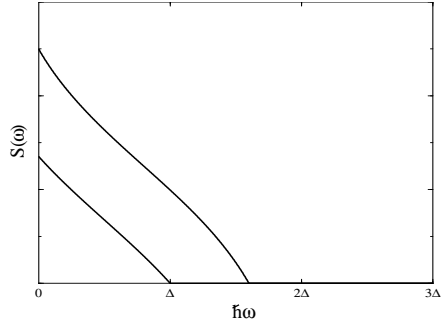

Fig. 10. Noise in an NS junction as a function of frequency, with a barrier with intermediate transparency ($Z = 1$). The applied bias is lower than the gap (from bottom to top, $eV = 0.5\Delta$ and $eV = 0.8\Delta$). The behavior for low biases is essentially linear, see Eq. (8.13).

### *8.3.3. Near and above gap regime*

In order to give predictions on the noise characteristic of a single NS junction when the voltage bias lies in the vicinity of the gap, it is necessary to specify a concrete model for the junction. A generic model was introduced by Blonder, Tinkham et Klapwijk (BTK) [78]: it has the advantage that the energy dependent scattering amplitudes can be derived using connection formulas from de Bogolubov-de Gennes equations. In particular, it allows to monitor the crossover from below gap to above gap regime. At the location of the junction, there is a superposition between a scalar delta function potential and a gap potential. The scalar potential acts on electrons or holes, and represents either a potential barrier or mimics the presence of disorder. The gap parameter is assumed to be a step function:

$$V_B(x) = V_B \delta(x) , \quad \text{with} \quad V_B = Z\hbar^2 k_F/m , \tag{8.19}$$
$$\Delta(x) = \Delta \Theta(x) . \tag{8.20}$$

$Z$ is the parameter which controls the transparency of the junction. $Z \gg 1$ corresponds to an opaque barrier.

In the preceding section, we made the assumption that the scattering amplitude had a weak dependence on energy. These assumptions can be tested with the BTK model. Fig. 10 displays the finite frequency noise for a bias which is half the gap and for a bias which approaches the gap, assuming an intermediate value of the transparency $Z = 1$. In the first case, we are very close to the ideal case of energy independent scattering amplitudes, while in the second case the linear dependency gets slightly distorted.

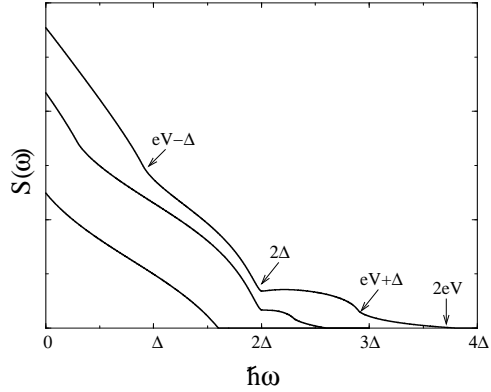

Fig. 11. Noise in an NS junction as a function of frequency, with a barrier with intermediate transparency ($Z = 1$), for biases below and above the gap $eV = 0.8\Delta$, $eV = 1.3\Delta$, $eV = 1.9\Delta$.

More interesting for the present model is the analysis of the noise when the bias voltage exceeds the gap. The scattering amplitudes were computed [78]. Inserting these in the energy integrals, the finite frequency noise now displays additional features (singularities) at $\omega = (eV - \Delta)/\hbar$, $\omega = 2\Delta/\hbar$, $\omega = (eV + \Delta)/\hbar$, in addition to the existing singularity at $\omega_J$. These frequency scales can be identified on the energy diagram of Fig. 12.

• Andreev reflection is still present. It implies a singularity at $\omega_J = 2eV/\hbar$.

• Electrons can be transmitted as electron-like quasiparticles, involving wave functions $\psi_{N,e} \sim \exp[-i(\mu_S + eV)t/\hbar]$ and $\psi_{S,e} \sim \exp[-i(\mu_S + \Delta)t/\hbar]$, thus a singularity at $\omega = (eV - \Delta)/\hbar$ (likewise for holes transmitted as hole-like quasiparticles).

• Electrons from the normal side can be transmitted as hole-like quasiparticles (Andreev transmission) with associated wave functions $\psi_{N,e} \sim \exp[-i(\mu_S + eV)t/\hbar]$ and $\psi_{S,h} \sim \exp[-i(\mu_S - \Delta)t/\hbar]$, giving a singularity at $\omega = (eV + \Delta)/\hbar$ (likewise for holes incident from the normal side being transmitted as electron-like quasiparticles in the superconductor).

• Andreev reflection also occurs for electron (hole) quasiparticles incident from the superconductors, reflected as holes (electrons) quasiparticles. Wave function are then $\psi_{S,e} \sim \exp[-i(\mu_S + \Delta)t/\hbar]$ and $\psi_{S,h} \sim \exp[-i(\mu_S - \Delta)t/\hbar]$, yielding a singularity at $2\omega = \Delta/\hbar$.

An extreme limit is the case where $\Delta \ll eV$. Transport is the fully dominated by the transfer of electrons and holes into quasiparticles, with a typical charge $e$ because Andreev reflection is not as important as before. We indeed recover (not shown) a noise characteristic similar to Fig. 8 (dashed line) with an abrupt change of slope at $\omega_J/2 = eV/\hbar$ characteristic of a normal metal junction. Finally, we mention that finite temperatures smear out all the structures in the finite frequency noise.

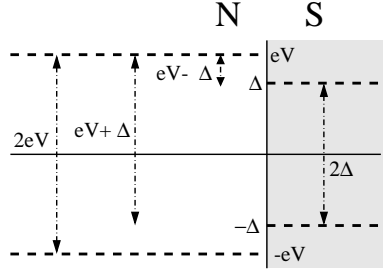

Fig. 12. Energy diagram of an NS junction when a bias is applied above the gap. Energy spacings in the frequency noise can be identified with the different processes: Andreev reflection (from either side), transmission of quasiparticles, and Andreev transmission.

### 8.4. *Hanbury-Brown and Twiss experiment with a superconducting source of electrons*

In this section, noise correlations are computed in a device which consists of two normal metal terminals (terminal 1 and 2, see Fig. 13) connected to an NS junction. The normal side of the NS junction is labelled 3, while the superconducting side is labelled 4. A junction playing the role of a beam splitter joins 1, 2 and 3. Let $c_{i\,e}^+$ ($c_{i\,e}^-$) label the state of an electron incident in (coming out from) terminal $i$. Likewise, incoming (outgoing) holes are labelled $c_{i\,h}^-$ ($c_{i\,h}^+$) (see Fig. 13). A scattering matrix $\mathcal{S}$ (describing both the beam splitter and the NS boundary) connects incoming states to outgoing states:

$$\begin{pmatrix} c_{1e}^- \\ c_{1h}^+ \\ c_{2e}^- \\ c_{2h}^+ \\ c_{4e}^- \\ c_{4h}^+ \end{pmatrix} = \mathcal{S} \begin{pmatrix} c_{1e}^+ \\ c_{1h}^- \\ c_{2e}^+ \\ c_{2h}^- \\ c_{4e}^+ \\ c_{4h}^- \end{pmatrix} . \qquad (8.21)$$

The noise correlations can be computed from the previous section, Eq. (8.12). In the limit of zero temperature, they can be shown to contain two contributions. The first one describes pure Andreev processes (involving one lead or both leads),

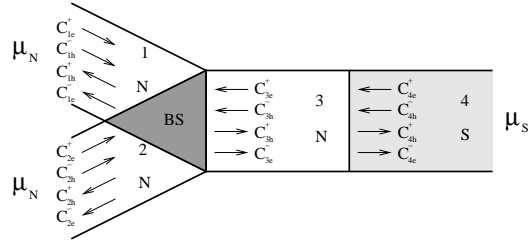

Fig. 13. 2 normal terminals (1 et 2) are connected by a beam splitter (BS), itself connected to a superconductor (4) via normal region (3).

while the second one involves above gap processes:

$$S_{12}(0) = \frac{2e^2}{h} \int_0^{eV} dE \Big[ \sum_{ij} \Big( s^*_{1iee} s_{1jeh} - s^*_{1ihe} s_{1jhh} \Big)$$
$$\times \Big( s^*_{2jeh} s_{2iee} - s^*_{2jhh} s_{2ihe} \Big)$$
$$+ \sum_{i\gamma} \Big( s^*_{1iee} s_{14e\gamma} - s^*_{1ihe} s_{14h\gamma} \Big)$$
$$\times \Big( s^*_{24e\gamma} s_{2iee} - s^*_{24h\gamma} s_{2ihe} \Big) \Big], \quad (8.22)$$

where $i, j = 1, 2$ and $\gamma = e, h$. However, the sign of correlations cannot be determined uniquely from Eq. (8.22). In the regime where electron-like and hole-like quasiparticles are transmitted in the normal terminals, one expects that the noise correlation will be negative because this situation is quite similar to the fermionic Hanbury-Brown and Twiss experiments. But what about the sub-gap regime ? Does it sustain positive or negative correlations ? Here a minimal model is chosen to describe the combination of the beam splitter and the NS junction, using a Fabry-Pérot analogy (Fig. 13).

*8.4.1. S–matrix for the beam splitter*
The electron part of the $S$ matrix for the beam splitter only connects the states:

$$\begin{pmatrix} c^-_{1e} \\ c^-_{2e} \\ c^-_{3e} \end{pmatrix} = S_{BS_e} \begin{pmatrix} c^+_{1e} \\ c^+_{2e} \\ c^+_{3e} \end{pmatrix}. \quad (8.23)$$

A simple version for this matrix $S_{BS_e}$, which is symmetric between 1 and 2, and whole elements are real, has been used profusely in mesoscopic physics transport

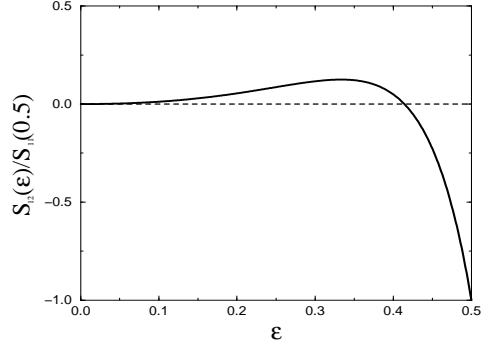

Fig. 14. Noise correlations between the two normal terminals (normalized to the autocorrelation noise in 1 or 2 for $\varepsilon = 1/2$) as a function of the beam splitter transparency $\varepsilon$. Correlations can either be positive or negative.

problems [79, 80]:

$$S_{BS_e} = \begin{pmatrix} a & b & \sqrt{\varepsilon} \\ b & a & \sqrt{\varepsilon} \\ \sqrt{\varepsilon} & \sqrt{\varepsilon} & -(a+b) \end{pmatrix},  \quad (8.24)$$

with this choice, the beam splitter $S$ matrix only depends on only one parameter. It unitarity imposes that $a = \left(\sqrt{1-2\varepsilon} - 1\right)/2$, $b = \left(\sqrt{1-2\varepsilon} + 1\right)/2$ where $\varepsilon \in [0, 1/2]$. $\varepsilon \ll 1/2$ means that the connection from 3 to 1 and 2 is opaque, while the opposite regime means a highly transparent connection to 1 and 2. Holes have similar scattering properties:

$$\begin{pmatrix} c_{1h}^+ \\ c_{2h}^+ \\ c_{3h}^+ \end{pmatrix} = S_{BS_h} \begin{pmatrix} c_{1h}^- \\ c_{2h}^- \\ c_{3h}^- \end{pmatrix}. \quad (8.25)$$

This choice for the beam splitter does not couple electron and holes: the superconductor boundary does that. The hole and the electron beam splitter S–matrix are related by $S_{BS_h}(E) = S^*_{BS_e}(-E)$ (in the absence of magnetic field). When one assumes that $\varepsilon$ does not depend on energy, both matrices are the same.

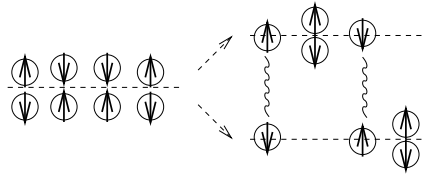

Fig. 15. (reverse) Andreev reflection emits pairs of electrons on the normal side. Correlations tend to be positive when the two electrons of the same pair go in opposite leads.

### 8.4.2. Sub-gap regime

When $eV \ll \Delta$, Andreev reflection is the only transport process at the NS boundary, and is described by the matrix [65]:

$$\begin{pmatrix} c_{3e}^+ \\ c_{3h}^- \end{pmatrix} = \begin{pmatrix} 0 & \gamma \\ \gamma & 0 \end{pmatrix} \begin{pmatrix} c_{3e}^- \\ c_{3h}^+ \end{pmatrix}, \qquad (8.26)$$

where $\gamma = e^{-i \arccos(E/\Delta)}$. At the same time, $s_{14\alpha\beta} = s_{24\alpha\beta} = 0$ ($\alpha\beta = e, h$). Combining the $S$ matrix of the beam splitter and that of the NS boundary, and defining $x = \sqrt{1 - 2\varepsilon}$, we obtain the elements of the $S$ matrix of the combined system:

$$s_{11ee} = s_{11hh} = s_{22ee} = s_{22hh} = \frac{(x-1)(1+\gamma^2 x)}{2(1-\gamma^2 x^2)}, \qquad (8.27)$$

$$s_{21ee} = s_{21hh} = s_{12ee} = s_{12hh} = \frac{(x+1)(1-\gamma^2 x)}{2(1-\gamma^2 x^2)}, \qquad (8.28)$$

$$\begin{aligned} s_{11eh} = s_{21eh} &= s_{12eh} = s_{22eh} = s_{11he} = s_{21he} = s_{12he} = s_{22he} \\ &= \frac{\gamma(1-x)(1+x)}{2(1-\gamma^2 x^2)} . \end{aligned} \qquad (8.29)$$

Since all energies are much smaller than the gap, we further simplify $\gamma \to -i$. Because the amplitudes do not depend on energy, the energy integral in Eq. (8.22) is performed and one obtains [70, 81]:

$$S_{12}(0) = \frac{2e^2}{h} eV \frac{\varepsilon^2}{2(1-\varepsilon)^4} \left(-\varepsilon^2 - 2\varepsilon + 1\right) . \qquad (8.30)$$

Noise correlations vanish at $\varepsilon = 0$, which corresponds to a two terminal junction between 1 et 2. $S_{12}$ also vanishes when $\varepsilon = \sqrt{2} - 1$. It is convenient to normalize $S_{12}$ (for arbitrary $\varepsilon$) with the autocorrelation noise in 1 or 2 computed at $\varepsilon = 1/2$ (see Fig. 14). Correlations are *positive* if $0 < \varepsilon < \sqrt{2} - 1$ and negative for $\sqrt{2} - 1 < \epsilon < 1/2$. Minimal negative correlations ($-1$) are reached when the connection to 1 and 2 is optimal: this is the signature of a purely fermionic system. Eq. (8.30) predicts a maximum in the positive correlations at $\epsilon = 1/3$.

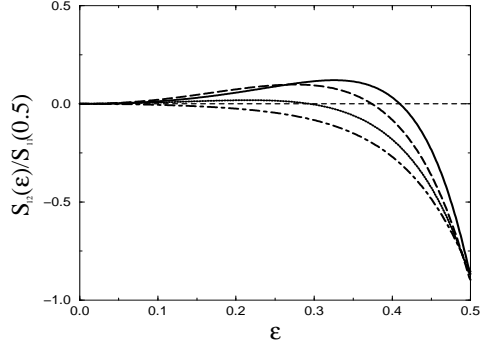

Fig. 16. Noise correlations between 1 and 2 (same normalization as before) as a function of $\varepsilon$. The NS junction is described within the BTK model, assuming a highly transparent barrier at the NS interface ($Z = 0.1$). Top to bottom, $eV/\Delta = 0.5, 0.95, 1.2, 1.8$.

Negative correlations correspond to Cooper pairs being distributed as a whole in the right or in the left arm.

Positive correlations have a simple interpretation. When a hole is reflected as an electron, this process can also be understood as a Cooper pair being emitted as two correlated electrons on the normal side [82]. It turns out that for an opaque beam splitter, the two electrons prefer to end up in opposite leads, giving a positive signal. This process is called the crossed Andreev process [83]. Other work, including full counting statistics approaches, describe in detail [84] why opaque barriers tend to favor positive correlations.

### 8.4.3. Near and above gap regime

The BTK model is chosen to describe the NS interface, in order to have the energy dependence of the scattering matrix elements. The integrals in Eq. (8.22) are computed numerically. As a first check, for a transparent interface ($Z = 0.1$) and a weak bias, one recovers the results of the previous section (see Fig. 16), except that the noise correlations do not quite reach the minimal value $-1$ for $\varepsilon = 1/2$, because of the presence of the barrier at the interface.

Note that noise correlations shift to negative values as one further increases the gap. For voltages beyond the gap, positive correlations disappear completely:

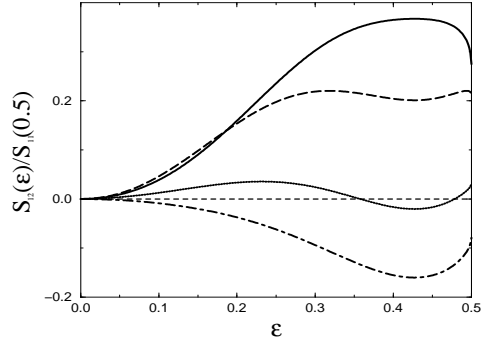

Fig. 17. Noise correlations between terminals 1 and 2 for a barrier with $Z = 1$ (same biases as in Fig. 16).

the system behaves like a normal metal junction.

What happens next if the NS interface has an appreciable Schottky barrier ? In Fig. 17, we consider a barrier with $Z = 1$, and the noise correlations change drastically: for small bias, the correlations are positive for all values of $\varepsilon$. This could be the regime where the experimental observation of positive correlation - in a fermionic system - could prove most likely: on the one hand an oxide barrier is needed, which reduces current flow at the junction, and thus a compromise has to be reached between signal detection and experimental conditions (opaque barrier) for observing the effect. For higher biases (par example $eV = 0.95\Delta$), one can monitor oscillations between positive and negative correlations. Yet, above the gap, the results are unchanged with respect to the high transparency case, as the correlations have a fermionic nature.

A number of approaches have shown the possibility of positive noise cross correlations in normal metal forks [84, 85].

## 9. Noise and entanglement

In quantum mechanics, a two-particle state is said to be entangled if a measurement on the state of one of the particles affects the outcome of the measurement of the state of the other particle. A celebrated example is the spin singlet:

$$\Psi_{12}\rangle = \frac{1}{\sqrt{2}} \left( |\uparrow_1, \downarrow_2\rangle| - |\downarrow_1, \uparrow_2\rangle \right) \tag{9.1}$$

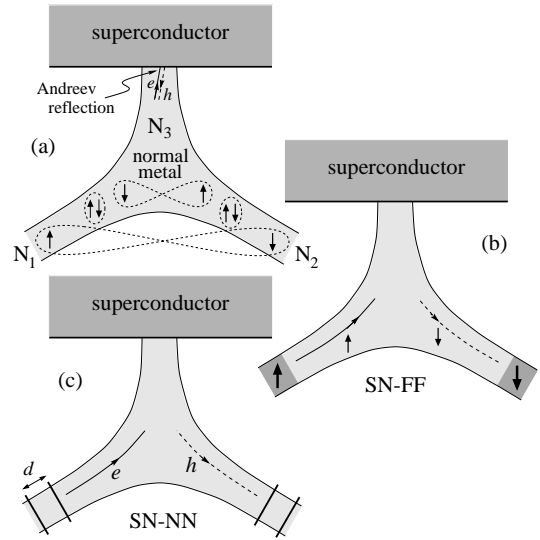

Fig. 18. Normal-metal–superconductor (NS) junction with normal-metal leads arranged in a fork geometry. (a) Without filters, entangled pairs of quasi-particles (Cooper pairs) injected in $N_3$ propagate into leads $N_1$ or $N_2$ either as a whole or one by one. The ferromagnetic filters in setup (b) separates the entangled spins, while the energy filters in (c) separate electron- and hole quasi-particles.

Entanglement is a crucial ingredient in most information processing schemes for quantum computation or for quantum communication. Here, we enquire whether entangled states of electrons can be generated in the vicinity of an s–wave superconductor, on the normal metal side.

## 9.1. Filtering spin/energy in superconducting forks

In the description of NS junction, one concluded that both positive and negative noise correlations were possible. Applying spin or energy filters to the normal arms 1 and 2 (Fig. 18), it is possible to generate positive correlations only [86]. For electrons emanating from a superconductors, it is possible to project either the spin or the energy with an appropriate filter, without perturbing the entanglement of the remaining degree of freedom (energy or spin). Energy filters, which are more appropriate towards a comparison with photon experiments, will have resonant energies symmetric above and below the superconductor chemical potential which serve to select electrons (holes) in leads $1(2)$. The positive correlation signal then reads:

$$S_{12}(0) = \frac{e^2}{h} \sum_{\zeta} \int_0^{e|V|} d\varepsilon \mathcal{T}_\zeta^A(\varepsilon)[1 - \mathcal{T}_\zeta^A(\varepsilon)] , \qquad (9.2)$$

where the index $\zeta = h, \sigma, 2$ $(h, -\sigma, 2)$, $(\sigma =\uparrow, \downarrow)$ identifies the incoming hole state for energy filters (positive energy electrons with arbitrary spin are injected in lead 1 here). $\zeta = h, \uparrow, 1$ $(h, \downarrow, 2)$ applies for spin filters (spin up electrons – with positive energy – emerging from the superconductor are selected in lead 1). $\mathcal{T}_\zeta^A$ is then the corresponding (reverse) crossed-Andreev reflection probability for each type of setup: the energy (spin) degree of freedom is frozen, while the spin (energy) degree of freedom is unspecified. $eV < 0$ insures that the constituent electrons of a Cooper pair from the superconductor are emitted into the leads without suffering from the Pauli exclusion principle. Moreover, because of such filters, the propagation of a Cooper pair as a whole in a given lead is prohibited. Note the similarity with the quantum noise suppression mentioned above. This is no accident: by adding constraints to our system, it has become a two terminal device, such that the noise correlations between the two arms are identical to the noise in one arm: $S_{11}(\omega = 0) = S_{12}(\omega = 0)$. The positive correlation and the perfect locking between the auto and cross correlations provide a serious symptom of entanglement. One can speculate that the wave function which describes the two electron state in the case of spin filters reads:

$$|\Phi_{\varepsilon,\sigma}^{\text{spin}}\rangle = \frac{1}{\sqrt{2}} \left(|\varepsilon, \sigma; -\varepsilon, -\sigma\rangle - |-\varepsilon, \sigma; \varepsilon, -\sigma\rangle\right) , \qquad (9.3)$$

where the first (second) argument in $|\phi_1; \phi_2\rangle$ refers to the quasi-particle state in lead 1 (2) evaluated behind the filters, $\varepsilon$ is the energy and $\sigma$ is a spin index. Note that by projecting the spin degrees of freedom in each lead, the spin entanglement is destroyed, but energy degrees of freedom are still entangled, and can help provide a measurement of quantum mechanical non-locality nevertheless. A measurement of energy $\varepsilon$ in lead 1 (with a quantum dot) projects the wave function so that the energy $-\varepsilon$ has to occur in lead 2. On the other hand, energy filters do preserve spin entanglement, and are appropriate to make a Bell test (see below). In this case the two electron wave function takes the form:

$$|\Phi_{\varepsilon,\sigma}^{\text{energy}}\rangle = \frac{1}{\sqrt{2}} \left(|\varepsilon, \sigma; -\varepsilon, -\sigma\rangle - |\varepsilon, -\sigma; -\varepsilon, \sigma\rangle\right) . \qquad (9.4)$$

Electrons emanating from the energy filters (coherent quantum dots) could be analyzed provided a measurement can be performed on the spin of the outgoing electrons with ferromagnetic leads.

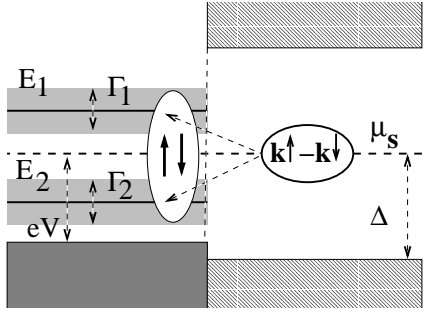

Fig. 19. Transfer of a Cooper pair on two quantum energy levels $E_{1,2}$ with a finite width $\Gamma_{1,2}$. The superconductor is located on the right hand side. The transfer of a Cooper pair gives an entangled state in the dots because it implies the creation and destruction of the same quasiparticle in the superconductor. The source drain voltage $eV$ for measuring noise correlations is indicated.

## 9.2. Tunneling approach to entanglement

We recall a perturbative argument which supports the claim that two electrons originating from the same Cooper pair are entangled. Consider a system composed of two quantum dots (energies $E_{1,2}$) next to a superconductor. An energy diagram is depicted in Fig. 19. The electron states in the superconductor are specified by the BCS wave function $|\Psi_{BCS}\rangle = \prod_k (u_k + v_k c^\dagger_{k\uparrow} c^\dagger_{-k\downarrow})|0\rangle$. Note that here one considers true electron creation operators, while previously we considered electron like and hole like quasiparticle operators. Tunneling to the dots is described by a single electron hopping Hamiltonian:

$$H_T = \sum_{kj\sigma} t_{jk} c^\dagger_{j\sigma} c_{k\sigma} + h.c. \,, \tag{9.5}$$

with $c^\dagger_{k\sigma}$ creates an electron with spin $\sigma$, and $j = 1, 2$. Now let us assume that the transfer Hamiltonian acts on a single Cooper pair.

Using the T-matrix to lowest (2nd) order, the wave function contribution of the two particle state with one electron in each dot and the superconductor back in its ground state reads:

$$\begin{aligned}|\delta\Psi_{12}\rangle &= H_T \frac{1}{i\eta - H_0} H_T |\Psi_{BCS}\rangle \\ &= [c^\dagger_{1\uparrow} c^\dagger_{2\downarrow} - c^\dagger_{1\downarrow} c^\dagger_{2\uparrow}] \sum_{jk} \frac{v_k u_k t_{1k} t_{2k}}{i\eta - E_k - E_j} |\Psi_{BCS}\rangle \,, \end{aligned} \tag{9.6}$$

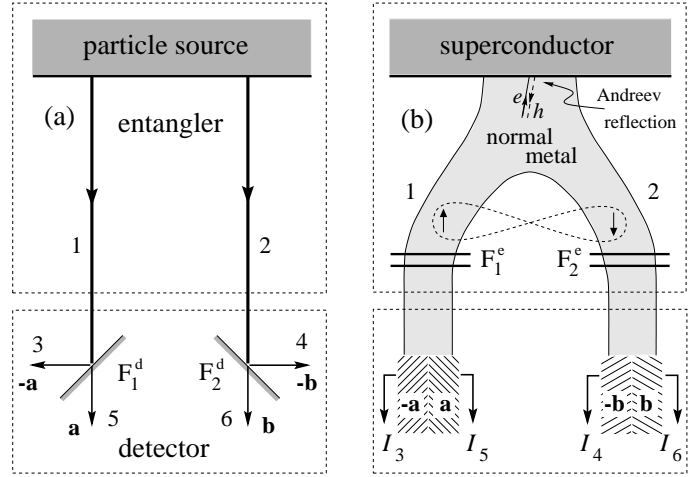

Fig. 20. a) Schematic setup for the measurement of Bell inequalities: a source emits particles into leads 1 and 2. The detector measures the correlation between beams labelled with odd and even numbers. Filters $F^d_{1(2)}$ select the spin: particles with polarization along the direction $\pm\mathbf{a}(\pm\mathbf{b})$ are transmitted through filter $F^d_{1(2)}$ into lead 5 and 3 (6 and 4). b) Solid state implementation, with superconducting source emitting Cooper pairs into the leads. Filters $F^e_{1,2}$ (e.g., Fabry-Perot double barrier structures or quantum dots) prevent Cooper pairs from entering a single lead. Ferromagnets with orientations $\pm\mathbf{a}$, $\pm\mathbf{b}$ play the role of the filters $F^d_{1(2)}$ in a); they are transparent for electrons with spin aligned along their magnetization axis.

where $E_k$ is the energy of a Bogolubov quasiparticle. The state of Eq. (9.6) has entangled spin degrees of freedom. This is clearly a result of the spin preserving tunneling Hamiltonian. Given the nature of the correlated electron state in the superconductor in terms of Cooper pairs, $H_T$ can only produce singlet states in the dots.

## 9.3. Bell inequalities with electrons

In photon experiments, entanglement is identified by a violation of Bell inequalities (BI) – which are obtained with a hidden variable theory. But in the case of photons, the BIs have been tested using photo-detectors measuring coincidence rates [87]. Counting quasi-particles one-by-one in coincidence measurements is difficult to achieve in solid-state systems where stationary currents and noise are the natural observables. Here, the BIs are re-formulated in terms of current-current cross-correlators (noise correlations) [88]. In order to derive Bell inequalities, we consider that a source provides two streams of particles (labeled 1 and 2) as in Fig. 20a injecting quasi-particles into two arms labelled by indices 1 and 2.

Filter $F_{1(2)}^d$ are transparent for electrons spin-polarized along the direction $\mathbf{a}(\mathbf{b})$. Assuming separability and locality [89] the density matrix for joint events in the leads $\alpha, \beta$ is chosen to be:

$$\rho = \int d\lambda f(\lambda) \rho_\alpha(\lambda) \otimes \rho_\beta(\lambda) , \tag{9.7}$$

where the lead index $\alpha$ is even and $\beta$ is odd (or vice-versa); the distribution function $f(\lambda)$ is positive. $\rho_\alpha(\lambda)$ are standard density matrices for a given lead, which are Hermitian. The total density matrix $\rho$ is the most general density matrix one can built for the source/detector system assuming no entanglement and only local correlations.

Consider an example of the solid-state analog of the Bell-device where the particle source is a superconductor in Fig. 20b. The chemical potential of the superconductor is larger than that of the leads, which means that electrons are flowing out of the superconductor. Two normal leads 1 and 2 are attached to it in a fork geometry [86, 90] and the filters $F_{1,2}^e$ enforce the energy splitting of the injected pairs. $F_{1,2}^d$-filters play the role of spin-selective beam-splitters in the detector. Quasi-particles injected into lead 1 and spin-polarized along the magnetization $\mathbf{a}$ enter the ferromagnet 5 and contribute to the current $I_5$, while quasi-particles with the opposite polarization contribute to the current $I_3$.

Consider the current operator $I_\alpha(t)$ in lead $\alpha = 1, \ldots, 6$ (see Fig. 20) and the associated particle number operator $N_\alpha(t, \tau) = \int_t^{t+\tau} I_\alpha(t')dt'$. Particle-number correlators are defined as:

$$\langle N_\alpha(t,\tau) N_\beta(t,\tau) \rangle_\rho = \int d\lambda f(\lambda) \langle N_\alpha(t,\tau) \rangle_\lambda \langle N_\beta(t,\tau) \rangle_\lambda , \tag{9.8}$$

with indices $\alpha/\beta$ odd/even or even/odd. The average $\langle N_\alpha(t,\tau) \rangle_\lambda$ depends on the state of the system in the interval $[t, t+\tau]$. An average over large time periods is introduced in addition to averaging over $\lambda$, e.g.,

$$\langle N_\alpha(\tau) N_\beta(\tau) \rangle \equiv \frac{1}{2T} \int_{-T}^{T} dt \langle N_\alpha(t,\tau) N_\beta(t,\tau) \rangle_\rho , \tag{9.9}$$

where $T/\tau \to \infty$ (a similar definition applies to $\langle N_\alpha(\tau) \rangle$). Particle number fluctuations are written as $\delta N_\alpha(t,\tau) \equiv N_\alpha(t,\tau) - \langle N_\alpha(\tau) \rangle$. Let $x, x', y, y', X, Y$ be real numbers such that:

$$|x/X|, |x'/X|, |y/Y|, |y'/Y| < 1 . \tag{9.10}$$

Then $-2XY \leq xy - xy' + x'y + x'y' \leq 2XY$. Define accordingly:

$$x = \langle N_5(t,\tau) \rangle_\lambda - \langle N_3(t,\tau) \rangle_\lambda, \tag{9.11}$$

$$x' = \langle N_{5'}(t,\tau)\rangle_\lambda - \langle N_{3'}(t,\tau)\rangle_\lambda, \tag{9.12}$$

$$y = \langle N_6(t,\tau)\rangle_\lambda - \langle N_4(t,\tau)\rangle_\lambda, \tag{9.13}$$

$$y' = \langle N_{6'}(t,\tau)\rangle_\lambda - \langle N_{4'}(t,\tau)\rangle_\lambda, \tag{9.14}$$

where the subscripts with a 'prime' indicate a different direction of spin-selection in the detector's filter (e.g., let $\mathbf{a}$ denote the direction of the electron spins in lead 5 ($-\mathbf{a}$ in lead 3), then the subscript $5'$ means that the electron spins in lead 5 are polarized along $\mathbf{a}'$ (along $-\mathbf{a}'$ in the lead 3). The quantities $X, Y$ are defined as

$$\begin{aligned}
X &= \langle N_5(t,\tau)\rangle_\lambda + \langle N_3(t,\tau)\rangle_\lambda \\
  &= \langle N_{5'}(t,\tau)\rangle_\lambda + \langle N_{3'}(t,\tau)\rangle_\lambda \\
  &= \langle N_1(t,\tau)\rangle_\lambda, \\
Y &= \langle N_6(t,\tau)\rangle_\lambda + \langle N_4(t,\tau)\rangle_\lambda \\
  &= \langle N_{6'}(t,\tau)\rangle_\lambda + \langle N_{4'}(t,\tau)\rangle_\lambda \\
  &= \langle N_2(t,\tau)\rangle_\lambda;
\end{aligned} \tag{9.15}$$

$$\tag{9.16}$$

The Bell inequality follows after appropriate averaging:

$$|F(\mathbf{a},\mathbf{b}) - F(\mathbf{a},\mathbf{b}') + F(\mathbf{a}',\mathbf{b}) + F(\mathbf{a}',\mathbf{b}')| \leq 2, \tag{9.17}$$

$$F(\mathbf{a},\mathbf{b}) = \frac{\langle [N_1(\mathbf{a},t) - N_1(-\mathbf{a},t)][N_2(\mathbf{b},t) - N_2(-\mathbf{b},t)]\rangle}{\langle [N_1(\mathbf{a},t) + N_1(-\mathbf{a},t)][N_2(\mathbf{b},t) + N_2(-\mathbf{b},t)]\rangle}, \tag{9.18}$$

with $\mathbf{a}, \mathbf{b}$ the polarizations of the filters $F_{1(2)}$ (electrons spin-polarized along $\mathbf{a}$ ($\mathbf{b}$) can go through filter $F_{1(2)}$ from lead 1(2) into lead 5(6)). This is the quantity we want to test, using a quantum mechanical theory of electron transport. Here it will be written in terms of noise correlators, as particle number correlators at equal time can be expressed in general as a function of the finite frequency noise cross-correlations. The correlator $\langle N_\alpha(\tau) N_\beta(\tau)\rangle$ includes both reducible and irreducible parts. The irreducible correlator $\langle \delta N_\alpha(\tau)\delta N_\beta(\tau)\rangle$ can be expressed through the shot noise power $S_{\alpha\beta}(\omega) = \int d\tau e^{i\omega\tau}\langle \delta I_\alpha(\tau)\delta I_\beta(0)\rangle$,

$$\langle \delta N_\alpha(\tau)\delta N_\beta(\tau)\rangle = \int_{-\infty}^{\infty}\frac{d\omega}{2\pi} S_{\alpha\beta}(\omega)\frac{4\sin^2(\omega\tau/2)}{\omega^2}. \tag{9.19}$$

In the limit of large times, $\sin^2(\omega\tau/2)/(\omega/2)^2 \to 2\pi\tau\delta(\omega)$, and therefore:

$$\langle N_\alpha(\tau) N_\beta(\tau)\rangle \approx \langle I_\alpha\rangle\langle I_\beta\rangle\tau^2 + \tau S_{\alpha\beta}(\omega = 0) \tag{9.20}$$

where $\langle I_\alpha\rangle$ is the average current in the lead $\alpha$ and $S_{\alpha\beta}$ denotes the shot noise. One then gets:

$$F(\mathbf{a},\mathbf{b}) = \frac{S_{56} - S_{54} - S_{36} + S_{34} + \Lambda_-}{S_{56} + S_{54} + S_{36} + S_{34} + \Lambda_+}, \tag{9.21}$$

where $\Lambda_\pm = \tau(\langle I_5 \rangle \pm \langle I_3 \rangle)(\langle I_6 \rangle \pm \langle I_4 \rangle)$ comes from the reducible part of the number correlators (the average number product). For a symmetric device, $\Lambda_- = 0$.

So far we have only provided a dictionary from the number correlator language used in optical measurements to the stationary quantities one encounters in nanophysics. We have provided absolutely no specific description of the physics which governs this beam splitter device. The test of the Bell inequality (9.17) requires information about the dependence of the noise on the mutual orientations of the magnetizations $\pm \mathbf{a}$ and $\pm \mathbf{b}$ of the ferromagnetic spin-filters. In the tunneling limit one finds the noise:

$$S_{\alpha\beta} = e \sin^2\left(\frac{\theta_{\alpha\beta}}{2}\right) \int_0^{|eV|} d\varepsilon \mathcal{T}^A(\varepsilon) , \qquad (9.22)$$

which integral also represents the current in a given lead (we have dropped the subscript in $\mathcal{T}^A(\varepsilon)$ assuming the two channels are symmetric). Here $\alpha = 3, 5$, $\beta = 4, 6$ or vice versa; $\theta_{\alpha\beta}$ denotes the angle between the magnetization of leads $\alpha$ and $\beta$, e.g., $\cos(\theta_{56}) = \mathbf{a} \cdot \mathbf{b}$, and $\cos(\theta_{54}) = \mathbf{a} \cdot (-\mathbf{b})$. Below, we need configurations with different settings $\mathbf{a}$ and $\mathbf{b}$ and we define the angle $\theta_{\mathbf{ab}} \equiv \theta_{56}$. $V$ is the bias of the superconductor.

The $\Lambda$-terms in Eq. (9.21) can be dropped if $\langle I_\alpha \rangle \tau \ll 1$, $\alpha = 3, \ldots, 6$, which corresponds to the assumption that only one Cooper pair is present on average. The resulting BIs Eqs. (9.17)-(9.21) then neither depend on $\tau$ nor on the average current but only on the shot-noise, and $F = -\cos(\theta_{\mathbf{ab}})$; the left hand side of Eq. (9.17) has a maximum when $\theta_{\mathbf{ab}} = \theta_{\mathbf{a'b}} = \theta_{\mathbf{a'b'}} = \pi/4$ and $\theta_{\mathbf{ab'}} = 3\theta_{\mathbf{ab}}$. With this choice of angles the BI Eq.(9.17) is *violated*, thus pointing to the nonlocal correlations between electrons in the leads 1,2 [see Fig. 20(b)].

If the filters have a width $\Gamma$ the current is of order $e\mathcal{T}^A\Gamma/\hbar$ and the condition for neglecting the reducible correlators becomes $\tau \ll \hbar/\Gamma\mathcal{T}^A$. On the other hand, in order to insure that no electron exchange between 1 and 2 one requires $\tau \ll \tau_{\mathrm{tr}}/\mathcal{T}^A$ ($\tau_{\mathrm{tr}}$ is the time of flight from detector 1 to 2). The conditions for BI violation require very small currents, because of the specification that one entangled pair at a time is in the system. Yet it is necessary to probe noise cross correlations of these same small currents. The noise experiments which we propose here are closely related to coincidence measurements in quantum optics. [87]

If we allow the filters to have a finite line width, which could reach the energy splitting of the pair, the violation of BI can still occur, although violation is not maximal. Moreover, when the source of electron is a normal source, our treatment has to be revised. The low frequency noise approximation to relate the number operators to the current operator breaks down at short times. Ref. [91]

shows in fact that entanglement can exist in ballistic forks. It is also possible to violate Bell inequalities if the normal source itself, composed of quantum dots as suggested in Ref. [92, 93], could generate entangled electron states as the result of electron-electron interactions.

Spin entanglement from superconducting source of electrons rely on the controlled fabrication of an entangler [90] or a superconducting beam-splitter with filters. Since the work of Ref. [88], other proposals for entanglement have been proposed, which avoid using spin. Electron-hole entanglement for a Hall bar with a point contact [94] exploits the fact that an electron, which can occupy either one of two edge channels, can either be reflected or transmitted. Unitary transformations between the two outgoing channels play the role of current measurement in arbitrary spin directions. Orbital entanglement using electrons emitted from two superconductors has been suggested in Ref. [95]. The electron-hole proposal of Ref. [94] has also been revisited using two distinct electron sources in a Hanbury-Brown and Twiss geometry together with additional beam splitters for detection, resulting in a a solid experimental proposal [96].

## 10. Noise in Luttinger liquids

In the previous sections, the noise was computed essentially in non-interacting systems. Granted, a superconductor depends crucially on the attractive interactions between electrons. Yet, the description of the NS boundary is like that of a normal mesoscopic conductor with electron and hole channels which get mixed. Now we want the noise in a system with repulsive interactions. There are many possibilities for doing that. One could consider transport through quantum dots where double occupancy of a dot costs a charging energy [38]. Alternatively one could consider to treat interactions in a mesoscopic conductor using perturbative many body techniques.

Instead we chose a situation where the interactions provide a genuine departure from single electron physics. The standard credo about interactions in condensed matter systems is the Fermi liquid picture. In two and three dimensions, it have been known for a long time that the quasiparticle picture holds: the elementary excitations of a system of interacting fermions resemble the original electrons. The excitations are named quasiparticles because their dynamics can be described in a similar manner as electrons, except for the fact that their mass is renormalized and that these quasiparticles have a finite lifetime. Perturbation theory, when done carefully in such systems, works rather well.

It is therefore more of a challenge to turn to the case of one dimensional systems where the Fermi liquid picture breaks down. Indeed, while in 2 and 3 dimensions the distribution function retains a step at the Fermi level, interactions in a one dimensional system render the distribution function continuous, with only a infinite derivative at the Fermi energy. But the more important feature of a one dimensional system is that the nature of the excitations changes drastically compared to its higher dimensional counterparts [97]. The excitations do not resemble electrons in any way: they consist of collective electron–hole excitations of the whole Fermi sea.

Luttinger liquid theory gives an account of the special properties of one-dimensional conductors [98]. For transport through an isolated impurity, the effect of interactions leads to a phase diagram [99]: in the presence of repulsive interactions, a weak impurity renders the wire insulating, while for positive interactions even a strong impurity is transparent.

The "easiest" type of Luttinger liquid arises on the boundaries of a sample which is put in the quantum Hall regime: a two dimensional electron gas (2DEG) under a high magnetic field. A classical description of such a system tells us that the electrons move along the edges, subject to the so called $\vec{E} \times \vec{B}$ drift. In mesoscopic physics, the electric field $\vec{E}$ comes from the confining electrostatic potential on the edges of the sample. Consider first a system of non-interacting electrons in the quantum regime (low temperature, high magnetic field). For an infinite sample, the quantum mechanical description gives the so called Landau levels, which are separated by the energy $\hbar\omega_c$, with $\omega_c = eB/mc$ the cyclotron frequency. Each landau level has a degeneracy $N_D = N_\phi$, with $N_\phi = BS/\phi_0$ the number of flux tubes which can be fitted a sample with area $S$. Given a magnetic field $B$ one defines the filling factor $\nu$ as the ratio between the total number of electrons and the number of flux tubes, or equivalently the fraction of the Landau levels which are filled. In the integer quantum Hall effect [100], the Landau energy spectrum allows to explain the quantification of the Hall resistance and the simultaneous vanishing of the longitudinal resistance when the magnetic field in varied (or, equivalently when the density of electrons is varied).

What happens when one considers confinement ? Landau levels bend upwards along the edges. So if one has adjusted the Fermi level of the system exactly between two landau levels, the highest populated states are precisely there. These are the quantum analogs of the classical skipping orbits. A very important feature is that they have a chiral character: they move only in one direction given one side of the sample, and in the opposite direction on the other side. This edge state description has allowed to explain in a rather intuitive manner the physics of the integer quantum Hall effect [33].

Interactions complicate things in a substantial manner, especially if one increases the magnetic field. When the lowest Landau level becomes partially filled, one reaches the fractional Hall effect regime [101, 102] (FQHE). The many body wave function for the electrons is such that it minimizes the effect of the interaction. The Hall resistance exhibits plateaus, $R_H = h/\nu e^2$ when the inverse of the filling factor is an odd integer. Spectacular effects follow. The excitation spectrum of this fractional quantum Hall fluid has a gap. The quasiparticles have a fractional charge, and if one exchanges the position of two such objects, the phase is neither 0 (bosons) nor $\pi$ (electrons) - quasiparticles have fractional statistics.

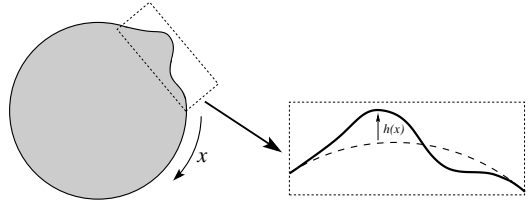

Fig. 21. Fractional quantum hole droplet. Excitations propagate along the edge.

Here, we want to know what happens at the edge. There are several arguments which justify the action which we shall use below. One of them relies on field theoretical arguments: starting from the fact that one is dealing with a gaped system, and effective action can be derived for the fluctuating electromagnetic field. If one now considers a finite fractional quantum Hall fluid with boundaries, one finds out that a boundary term must be added in order to preserve gauge invariance. This term turn out to generate the dynamics of the edge excitations. Here however, I will use a more intuitive argument to describe the edges called the hydrodynamic approach.

*10.1. Edge states in the fractional quantum Hall effect*

The Hamiltonian which describes the edge modes at the edge is simply an electrostatic term: [103]

$$H = \frac{1}{2} \int_0^L V(x) e \rho(x) dx , \quad (10.1)$$

with $x$ a curvilinear coordinate along the edge and where $V(x)$ is the confining potential. This potential is related to the confining electric field as $E = -\nabla V \sim \partial_y V$, with $y$ the coordinate perpendicular to the edge. $E$ and $B$ are related because the drift velocity is given by $|\vec{v}| = c|\vec{E} \times \vec{B}|/B^2$. The electrostatic potential can then be expressed in terms of the lateral displacement of the quantum Hall fluid $h(x)$, which is also expressed in terms of the linear charge density $\rho(x)$:

$$V(x) = Eh(x) = (vB/c)(\rho(x)/n_s) . \quad (10.2)$$

Inserting this in the Hamiltonian, we find the remarkable property that the Hamiltonian is quadratic in the density. At this point it is useful to use the definition of the flux quantum $\phi_0 = hc/e$ in order to eliminate the 2D electron density $n_s$ from the problem in favor of the filling factor $\nu$.

$$H = \frac{1}{2}\frac{hv}{\nu}\int_0^L \rho^2(x)\,dx. \tag{10.3}$$

So far one has used a purely classical argument. In order to obtain a quantum mechanical description, we need to impose quantification rules. First, it is convenient to transform this Hamiltonian into Fourier space using:

$$\rho(x) = \frac{1}{\sqrt{L}}\sum_k e^{-ikx}\rho_k, \tag{10.4}$$

$$H = \frac{1}{2}\frac{vh}{\nu}\sum_k \rho_k \rho_{-k}. \tag{10.5}$$

Quantification requires first to identify a set of canonical conjugate variables $q_k$ and $p_k$ which satisfy Hamilton's equations. Identifying $q_k = \rho_k$, one obtains

$$\dot{p}_k = -\frac{\partial H}{\partial \rho_k} = -\frac{1}{2}\frac{vh}{\nu}2\rho_{-k}. \tag{10.6}$$

The continuity equation for this chiral density reads $\dot{\rho}_{-k} = -vik\rho_{-k}$. Integrating over time one thus get the canonical conjugate:

$$p_k = -i\frac{h}{\nu}\frac{\rho_{-k}}{k}. \tag{10.7}$$

Quantification is achieved by imposing the commutation relation

$$[q_k, p_{k'}] = i\hbar\,\delta_{kk'}. \tag{10.8}$$

Note this is exactly the same procedure as one uses for phonons in conventional condensed matter physics. Replacing $p_k$ by its expression in Eq. (10.7) one gets the Kac-Moody commutation relations:

$$[\rho_k, \rho_{k'}] = -\frac{\nu k}{2\pi}\delta_{k'\,-k}. \tag{10.9}$$

Computing the commutator of the Hamiltonian with the density, one gets

$$[H, \rho_{k'}] = v\hbar k'\rho_{k'}, \tag{10.10}$$

and one sees that the Heisenberg evolution equation $i\hbar \dot\rho_k = [H, \rho_k]$ give the continuity equation. We now turn to the definition of the electron operator. Because $\rho(x)$ is the charge density, we expect the electron creation operator to satisfy

$$[\rho(x'), \psi^\dagger(x)] = \delta(x-x')\psi^\dagger(x) , \qquad (10.11)$$

which is equivalent to saying that the measurement of the electronic density on a state where $\psi^\dagger(x)$ is acting tells us than an electron has been added. This is the same commutation relation one uses in the derivation of the raising and lowering operators. For later purposes, the Luttinger bosonic field is introduced:

$$\phi(x) = \frac{\pi}{\sqrt{\nu}} \frac{1}{\sqrt{L}} \sum_k i \frac{e^{-a|k|/2}}{k} e^{-ikx} \rho_k . \qquad (10.12)$$

Where the factor $a$ takes the meaning of a spatial cutoff similar to that used in non-chiral Luttinger liquids. Here it insures the convergence of the integral. What is important about this definition is that the derivative of $\phi$ is proportional to the density:

$$\frac{\partial \phi}{\partial x} = \frac{\pi}{\sqrt{\nu}} \rho(x) . \qquad (10.13)$$

This allows to re-express the Hamiltonian in terms of $\phi$ [104]:

$$H = \frac{\hbar v}{\pi} \int_0^L (\frac{\partial \phi}{\partial x})^2 dx . \qquad (10.14)$$

The form of the electron operators is found by an analogy with the properties of canonical conjugate variables $p(x)$ and $q(x)$:

$$[p(x), q(x')] = -i\delta(x-x') \to [p(x), e^{iq(x')}] = \delta(x-x')e^{iq(x)} . \qquad (10.15)$$

Next, one can identify $p$ as $\rho$ and $q$ as $\phi/\sqrt{\nu}$. Using the Kac-Moody commutation relations: $[\rho(x'), \nu^{-1/2}\phi(x)] = -i\delta(x-x')$ Comparing with the relation (10.15) and the definition (10.11), the annihilation operator takes the form:

$$\psi(x) = \frac{1}{\sqrt{2\pi a}} e^{-ikx} e^{i\frac{1}{\sqrt{\nu}}\phi(x)} , \qquad (10.16)$$

with $e^{ikx}$ giving the phase accumulated along the edge. This operator obviously depends on the filling factor. Fermion operators are known to anti-commute, so what are the constraints on this filling factor in order to insure anti-commutation relations $\{\psi(x), \psi(x')\} = 0$. The anti-commutator can be computed using the Baker-Campbell-Hausdorff formula: $e^A e^B = e^{A+B-\frac{[A,B]}{2}}$ which is only true of the commutator is a $c$-number. One thus needs the commutation relation of the bosonic field:

$$[\phi(x), \phi(x')] = -i\pi sgn(x - x') . \tag{10.17}$$

The two products of fermionic operators is then:

$$\psi(x)\psi(x') = \frac{1}{(2\pi a)^2} e^{ik(x+x')} e^{i\frac{1}{\sqrt{\nu}}\phi(x)+\phi(x')} e^{\frac{i\pi}{2\nu} sgn(x-x')} . \tag{10.18}$$

So one concludes that:

$$\psi(x)\psi(x') = e^{\pm i\frac{\pi}{\nu}} \psi(x')\psi(x) . \tag{10.19}$$

In order to insure anti-commutation relations, one need to set $\nu = 1/m$ with $m$ an odd integer. This conclusion is consistent with the assumption that in the bulk, one is dealing with a fractional quantum Hall fluid.

In order to obtain information on the dynamics of electrons (or of fractional quasiparticles), one needs to specify the bosonic Green's function. It is thus convenient to derive the action for this bosonic field. The Lagrangian is obtained from a Legendre transformation on the Hamiltonian, taking as canonical conjugate variables $\phi(x)$ and $-i(\hbar/\pi)\partial_x \phi$ in accordance with the Kac-Moody relations. The Euclidean action then reads:

$$S_E = -\frac{\hbar}{\pi} \int d\tau \int dx \left[\partial_x \phi \left(v \partial_x + i \partial_\tau \right) \phi \right] . \tag{10.20}$$

The operator which is implicit in this quadratic action allows to define the Green's function $G(x, \tau) = \langle T_\tau \phi(x, \tau) \phi(0, 0) \rangle$ – the correlation function of the bosonic field. This Green's function is defined by the differential equation:

$$(i\partial_\tau + v\partial_x)\partial_x G(x, \tau) = 2\pi \delta(x) \delta(\tau) . \tag{10.21}$$

The solution of this equation is obtained by setting $-\partial_x G = f$, and using the complex variables $z = x/v + i\tau$ et $\bar{z} = x/v - i\tau$. The equation for $f$ becomes:

$$\partial_{\bar{z}} f = v\pi \delta(x) \delta(\tau) . \tag{10.22}$$

From two dimensional electrostatics, it can be justified that $f(z) = 1/z$. Yet, one is dealing here with the thermal Green's function, which must be a periodic function of $\tau$ with period $\beta$, so a periodic extension of $f(z)$ is given:

$$f(z + i\beta) = f(z) = \frac{1}{z} + \sum_{n \neq 0} \frac{1}{z - in\beta} = \frac{\pi}{v\beta} \coth\left(\pi \frac{z}{\beta}\right) . \tag{10.23}$$

The thermal Green's function is subsequently obtained by integrating over $x$:

$$G(x,\tau) = -\ln\left[\sinh\left(\pi\frac{x/v + i\tau}{\beta}\right)\right] . \tag{10.24}$$

*10.2. Transport between two quantum Hall edges*

The noise has been computed for a single point contact [103, 105, 106]. It is typically achieved by placing metallic gates on top of the 2DEG and applying a potential to deplete the electron gas underneath the gates. By varying the gate potential, one can switch from a weak backscattering situation, where the Hall liquid remains in one piece, to a strong backscattering situation where the Hall fluid is split into two. In the former case, the entities which tunnel are edge quasiparticle excitations. In the latter case, the general convention is to say that in between the two fluids, only electrons can tunnel, because nothing can "dress" these electrons into strange quasiparticles like in the previous case. Here we will focus mostly on the weak backscattering case, because this is the situation where the physics of FQHE quasiparticles is most obvious. Anyway, the description of the strong backscattering case can be readily obtained using a duality transformation.

The tunneling Hamiltonian describing the coupling between the two edges $L$ and $R$ is like a tight binding term, where for convenience we use a compact notation [57] to describe the two hermitian conjugate parts:

$$H_{int} = \sum_{\varepsilon=\pm}\left[\Gamma_0\Psi_R^\dagger\Psi_L\right]^\varepsilon \text{ with } \begin{cases} \left[\Gamma_0\Psi_R^\dagger\Psi_L\right]^+ = \Gamma_0\Psi_R^\dagger\Psi_L \\ \left[\Gamma_0\Psi_R^\dagger\Psi_L\right]^- = \Gamma_0^*\Psi_L^\dagger\Psi_R \end{cases} \tag{10.25}$$

Where the quasiparticle operators have the form:

$$\Psi_{R(L)}(t) = \frac{M_{R(L)}}{\sqrt{2\pi a}}e^{i\sqrt{\nu}\phi_{R(L)}(t)} , \tag{10.26}$$

where the spatial cutoff is defined as $a = v\tau_0$, where $\tau_0$ is the temporal cutoff. $M_{R(L)}$ is a Klein factor, which insures the proper statistical properties. Nevertheless, for problems involving only two edges, it turns out to be irrelevant and it will be omitted. The quasiparticles are, in a sense, the $1/\nu$ root of the electron operators.

In scattering theory, the bias voltage was included by choosing appropriately the chemical potentials of the reservoirs, that is in non-equilibrium theory language, in the energy representation of the Green's functions for the leads. Here it is more difficult, because the Green's function are defined in real time, and the density of states of FQHE quasiparticles diverges at the Fermi level. The trick is to proceed with a gauge transformation. Starting from a gauge where the electric field is solely described by the scalar potential, $A = 0$, we proceed to a gauge transformation such that the new scalar potential is zero:

$$\begin{cases} V' &= V - \frac{1}{c}\partial_t \chi &= 0 \\ \vec{A}' &= \vec{A} + \nabla \chi &\neq 0 \\ \Psi'_i &= e^{i\frac{e^*\chi_i}{\hbar c}}\Psi_i &, e^* = \nu e \end{cases} \quad \text{so } \nabla \chi = \vec{A}' \quad (10.27)$$

For a constant potential along the edges, the gauge function $\chi$ depends only on time, and $\Delta \chi = \int_L^R \vec{A}'.\vec{dl} = cV_0 t$. Because we are dealing with quasiparticle transfer we *anticipate* that the quasiparticle charge is $e^* = \nu e$. Upon gauge transforming the quasiparticle operators, the tunneling amplitude becomes:

$$\Gamma_0 \to \Gamma_0 e^{i\omega_0 t} , \quad (10.28)$$

where $\omega_0 = e^* V_0$. From this expression, the backscattering current operator is derived from the Heisenberg equation of motion for the density, or alternatively by calculating $I_B(t) = -c\partial H_B(t)/\partial \chi(t)$:

$$I_B(t) = ie^* \Gamma_0 \sum_\varepsilon \varepsilon e^{i\varepsilon\omega_0 t} [\Psi_R^\dagger(t)\Psi_L(t)]^{(\varepsilon)} . \quad (10.29)$$

### 10.3. Keldysh digest for tunneling

In many body physics, it is convenient to work with a Wick's theorem (or one of its generalizations) in order to compute products of fermion and bosons operators. It is encountered when one considered averages of Heisenberg operators ordered in time, and one is faced with the problem of translating this into interaction representation products. The problem with the Heisenberg representation is that the operators contains the "difficult" part (the interaction part) of the Hamiltonian. Consider the ground state average of a time-ordered product of Heisenberg operators:

$$\langle A_H(t_0) B_H(t_1) C_H(t_2) D_H(t_3) \ldots \rangle \text{ with } t_0 > t_1 > t_2 > t_3 > \ldots \quad (10.30)$$

when translating to the interaction representation, the evolution operator reads:

$$S(t,t') = \hat{T} exp\left\{-i \int_{t'}^{t} dt'' H_{int}(t'') \right\} , \quad (10.31)$$

all operators such as $H_{int}$ become $e^{iH_0 t} H_{int} e^{-iH_0 t}$ [107] in this language. The product of ordered operators then becomes:

$$\langle S(-\infty, +\infty) \hat{T}(A_I(t_0) B_I(t_1) C_I(t_2) D_I(t_3) \ldots S(+\infty, -\infty)) \rangle , \quad (10.32)$$

where $\hat{T}$ is the time ordering operator. When the system is at zero temperature or in equilibrium, the ground state (or thermal) expectation of this S–matrix is just a phase factor, because one assumes that the perturbation is turned on adiabatically. This means that $\langle S(+\infty, -\infty) \rangle = e^{i\gamma}$. One is therefore left with a T–product which is easily computed with the help of Wick's theorem.

However, if the system is off equilibrium, one cannot a priori used Wick theorem to compute the average: the S-matrix in front of the T–product spoils everything because particles are being transfered from one reservoir to the other, and the ground state at $t = +\infty$ does not look like anything like the the ground state at $t = -\infty$ (both are no more related by a phase factor). To remedy this problem, Keldysh proposed to invent a new contour, which goes from $t = -\infty$ to $t = +\infty$ and back to $t = -\infty$, and a corresponding new time ordering operator $\hat{T}_K$. Because times on the lower contour are "larger" than times on the upper contour, the product of operators can be written as:

$$\langle \hat{T}_K(A_I(t_0)B_I(t_1)C_I(t_2)D_I(t_3)\ldots S(-\infty, -\infty)) \rangle \,, \tag{10.33}$$

where the integral over the Keldysh contour $K$ goes from $-\infty$ to $+\infty$ and then back to $-\infty$. Note that in general, the times appearing in the operator product $A_I(t_0)B_I(t_1)C_I(t_2)D_I(t_3)$ can be located either on the upper or on the lower contour. The Green's function associated with the two-branches Keldysh contour is therefore a $2 \times 2$ matrix:

$$\begin{aligned}
\tilde{G}(t - t') &= \begin{pmatrix} \tilde{G}^{++}(t - t') & \tilde{G}^{+-}(t - t') \\ \tilde{G}^{-+}(t - t') & \tilde{G}^{--}(t - t') \end{pmatrix} \\
&= \begin{pmatrix} \tilde{G}^{-+}(|t - t'|) & \tilde{G}^{-+}(t' - t) \\ \tilde{G}^{-+}(t - t') & \tilde{G}^{-+}(-|t - t'|) \end{pmatrix} \,,
\end{aligned} \tag{10.34}$$

where $\tilde{G}^{-+}(t)$ can be computed from the thermal Green's function using a Wick rotation. Often one redefines the Green's function by subtracting to the initial Green's function its equal time arguments (see below).

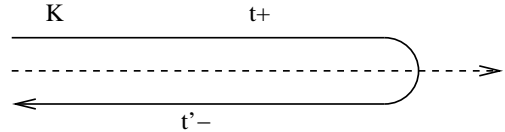

Fig. 22. The two-branches of the Keldysh contour

*10.4. Backscattering current*

For the calculation of an operator which involves a single time argument, it does not matter on which branch of the Keldysh contour we assign the time. We therefore chose a symmetric combination:

$$\langle I_B(t)\rangle = \frac{1}{2}\sum_\eta \langle \hat{T}_K\{I_B(t^\eta)e^{-i\int_K dt_1 H_B(t_1)}\}\rangle \,. \tag{10.35}$$

To lowest order in the tunnel amplitude $\Gamma_0$, we have:

$$\begin{aligned}\langle I_B(t)\rangle &= \frac{e^*\Gamma_0^2}{2}\sum_{\eta\eta_1\varepsilon\varepsilon_1}\varepsilon\eta_1\int_{-\infty}^{+\infty}dt_1 e^{i\varepsilon\omega_0 t+i\varepsilon_1\omega_0 t_1}\\ &\times \langle \hat{T}_K\{[\Psi_R^\dagger(t^\eta)\Psi_L(t^\eta)]^{(\varepsilon)}[\Psi_R^\dagger(t_1^{\eta_1})\Psi_L(t_1^{\eta_1})]^{(\varepsilon_1)}\}\rangle\,.\end{aligned} \tag{10.36}$$

The correlator is different from zero only when $\varepsilon_1 = -\varepsilon$. This amounts to saying that quasiparticles are conserved in the tunneling process. The sum over $\varepsilon$ gives, after inserting the chiral bosonic field $\phi_{R(L)}$:

$$\begin{aligned}\langle I_B(t)\rangle &= \frac{e^*\Gamma_0^2 M_R^2 M_L^2}{8\pi^2 a^2}\sum_{\eta\eta_1}\eta_1\int_{-\infty}^{+\infty}dt_1\\ &\times \Big(e^{i\omega_0(t-t_1)}\langle \hat{T}_K\{e^{-i\sqrt{\nu}\phi_R(t^\eta)}e^{i\sqrt{\nu}\phi_L(t^\eta)}e^{-i\sqrt{\nu}\phi_L(t_1^{\eta_1})}e^{i\sqrt{\nu}\phi_R(t_1^{\eta_1})}\}\rangle\\ &\quad - e^{-i\omega_0(t-t_1)}\langle \hat{T}_K\{e^{-i\sqrt{\nu}\phi_L(t^\eta)}e^{i\sqrt{\nu}\phi_R(t^\eta)}e^{-i\sqrt{\nu}\phi_R(t_1^{\eta_1})}e^{i\sqrt{\nu}\phi_L(t_1^{\eta_1})}\}\rangle\Big)\,.\end{aligned} \tag{10.37}$$

We use $M_{R(L)}^2 = 1$, and introduce the chiral Green's function of the bosonic field $\tilde{G}^{\eta\eta'}(t-t') = \langle \hat{T}_K\{\phi_{R(L)}(t^\eta)\phi_{R(L)}(t'^{\eta'})\}\rangle - \frac{1}{2}\langle \hat{T}_K\{\phi_{R(L)}(t^\eta)^2\}\rangle - \frac{1}{2}\langle \hat{T}_K\{\phi_{R(L)}(t'^{\eta'})^2\}\rangle$ which does not depend on the chirality $R(L)$. We obtain the expression for the backscattering current:

$$\langle I_B(t)\rangle = \frac{ie^*\Gamma_0^2}{4\pi^2 a^2}\sum_{\eta\eta_1}\eta_1\int_{-\infty}^{+\infty}dt_1 \sin(\omega_0(t-t_1))e^{2\nu\tilde{G}^{\eta\eta_1}(t-t_1)} \,. \tag{10.38}$$

Because the Green's function $\tilde{G}^{\eta\eta}$ is an even function (see Eq. (10.34)), the contributions $\eta = \eta_1$ vanish. We perform the change of variables: $\tau = t - t_1$ with $d\tau = -dt_1$, then:

$$\langle I_B(t)\rangle = -\frac{ie^*\Gamma_0^2}{4\pi^2 a^2}\sum_\eta \eta\int_{-\infty}^{+\infty}d\tau \sin(\omega_0\tau)e^{2\nu\tilde{G}^{\eta,-\eta}(\tau)} \,. \tag{10.39}$$

At zero temperature, the off-diagonal Keldysh Green's function is $\tilde{G}^{\eta-\eta}(\tau) = -\ln(1 - i\eta v_F \tau/a)$. Thus, we have:

$$\langle I_B(t) \rangle = -\frac{ie^*\Gamma_0^2}{4\pi^2 a^2} \sum_\eta \eta \int_{-\infty}^{+\infty} d\tau \frac{\sin(\omega_0 \tau)}{(1 - i\eta\tau v_F/a)^{2\nu}} \ . \tag{10.40}$$

Performing the integration, we obtain the final result:

$$\langle I_B(t) \rangle = \frac{e^*\Gamma_0^2}{2\pi a^2 \mathbf{\Gamma}(2\nu)} \left(\frac{a}{v_F}\right)^{2\nu} \text{sgn}(\omega_0)|\omega_0|^{2\nu-1} \ , , \tag{10.41}$$

where $\mathbf{\Gamma}$ is the gamma function.

On the other hand, at finite temperatures, the Green's function is given by:

$$\tilde{G}^{\eta-\eta}(\tau) = -\ln\left(\sinh\left(\frac{\pi}{\beta}(\eta\tau + i\tau_0)\right) / \sinh\left(\frac{i\pi\tau_0}{\beta}\right)\right) \ , \tag{10.42}$$

where $\tau_0 = a/v_F$. The average current is then given by the integral:

$$\langle I_B(t) \rangle = -\frac{ie^*\Gamma_0^2}{4\pi^2 a^2} \sum_\eta \eta \int_{-\infty}^{+\infty} d\tau \sin(\omega_0\tau) \left(\frac{\sinh\left(\frac{i\pi\tau_0}{\beta}\right)}{\sinh\left(\frac{\pi}{\beta}(\eta\tau + i\tau_0)\right)}\right)^{2\nu} \ . \tag{10.43}$$

The change of variables $t = -\tau - i\eta\tau_0 + i\eta\beta/2$ with $dt = -d\tau$ is operated. The time integral now runs in the complex plane from $-\infty - i\eta\tau_0 + i\eta\beta/2$ to $+\infty - i\eta\tau_0 + i\eta\beta/2$. We can bring is back to $-\infty$ to $+\infty$ provided that there are no poles in the integrand, encountered when changing the contour. The poles are located at integer values of $i\pi$ and $i\pi/2$: for this reason the presence of the cutoff is crucial as depending on the sign of $\eta$, one is always allowed to deform the contour to the real axis. The integral becomes:

$$\langle I_B(t) \rangle = \frac{e^*\Gamma_0^2}{2\pi^2 a^2} \left(\frac{\pi\tau_0}{\beta}\right)^{2\nu} \sinh\left(\frac{\omega_0\beta}{2}\right) \int_{-\infty}^{+\infty} dt \frac{\cos(\omega_0 t)}{\cosh^{2\nu}\left(\frac{\pi t}{\beta}\right)} \ . \tag{10.44}$$

The integral can be computed analytically:

$$\langle I_B(t) \rangle = \frac{e^*\Gamma_0^2}{2\pi^2 a^2 \mathbf{\Gamma}(2\nu)} \left(\frac{a}{v_F}\right)^{2\nu} \left(\frac{2\pi}{\beta}\right)^{2\nu-1}$$
$$\times \sinh\left(\frac{\omega_0\beta}{2}\right) \left|\mathbf{\Gamma}\left(\nu + i\frac{\omega_0\beta}{2\pi}\right)\right|^2 \ . \tag{10.45}$$

*10.5. Poissonian noise in the quantum Hall effect*

Using the symmetric combination of the noise correlators (we are interested in zero frequency noise):

$$\begin{aligned} S(t,t') &= \langle I_B(t)I_B(t')\rangle + \langle I_B(t')I_B(t)\rangle - 2\langle I_B(t)\rangle\langle I_B(t')\rangle \\ &= \sum_\eta \langle \hat{T}_K\{I_B(t^\eta)I_B(t'^{-\eta})e^{-i\int_K dt_1 H_B(t_1)}\}\rangle - 2\langle I_B\rangle^2 \,, \end{aligned}$$
(10.46)

to lowest order in the tunnel amplitude $\Gamma_0$, it is not even necessary to expand the Keldysh evolution operator because the current itself contains $\Gamma_0$.

$$\begin{aligned} S(t,t') &= -e^{*2}\Gamma_0^2 \sum_{\eta\varepsilon\varepsilon'} \varepsilon\varepsilon' e^{i\varepsilon\omega_0 t} e^{i\varepsilon'\omega_0 t'} \\ &\quad \times \langle \hat{T}_K\{[\Psi_R^\dagger(t^\eta)\Psi_L(t^\eta)]^{(\varepsilon)}[\Psi_R^\dagger(t'^{-\eta})\Psi_L(t'^{-\eta})]^{(\varepsilon')}\}\rangle \,. \end{aligned}$$
(10.47)

The correlator is different from zero only when $\varepsilon' = -\varepsilon$ such correlators have already been calculated for the current:

$$S(t,t') = \frac{e^{*2}\Gamma_0^2}{2\pi^2 a^2} \sum_\eta \cos(\omega_0(t-t'))e^{2\nu G^{\eta-\eta}(t-t')} = S(t-t') \,. \qquad (10.48)$$

From this expression, the Fourier transform at zero frequency is computed, first at zero temperature:

$$\begin{aligned} S(\omega=0) &= \frac{e^{*2}\Gamma_0^2}{2\pi^2 a^2} \sum_\eta \int_{-\infty}^{+\infty} dt\, \cos(\omega_0 t) e^{2\nu G^{\eta-\eta}(t)} \\ &= \frac{e^{*2}\Gamma_0^2}{2\pi a^2 \mathbf{\Gamma}(2\nu)} \left(\frac{a}{v_F}\right)^{2\nu} |\omega_0|^{2\nu-1} \,. \end{aligned}$$
(10.49)

The Schottky relation applies, with a fractional charge $e^* = \nu e$:

$$S(\omega=0) = 2e^*|\langle I_B(t)\rangle| \,. \qquad (10.50)$$

At finite temperature, the noise is given by the integral:

$$S(\omega=0) = \frac{e^{*2}\Gamma_0^2}{2\pi^2 a^2} \sum_\eta \int_{-\infty}^{+\infty} d\tau \cos(\omega_0 \tau) \left(\frac{\sinh\left(\frac{i\pi\tau_0}{\beta}\right)}{\sinh\left(\frac{\pi}{\beta}(\eta\tau+i\tau_0)\right)}\right)^{2\nu} \,. \qquad (10.51)$$

Performing the same change of variables as for the current, this leads to:

$$S(\omega = 0) = \frac{e^{*2}\Gamma_0^2}{\pi^2 a^2}\left(\frac{\pi\tau_0}{\beta}\right)^{2\nu}\cosh\left(\frac{\omega_0\beta}{2}\right)\int_{-\infty}^{+\infty}dt\frac{\cos(\omega_0 t)}{\cosh^{2\nu}\left(\frac{\pi t}{\beta}\right)}. \quad (10.52)$$

Performing the integral:

$$S(\omega = 0) = \frac{e^{*2}\Gamma_0^2}{\pi^2 a^2 \mathbf{\Gamma}(2\nu)}\left(\frac{a}{v_F}\right)^{2\nu}\left(\frac{2\pi}{\beta}\right)^{2\nu-1}$$
$$\times \cosh\left(\frac{\omega_0\beta}{2}\right)\left|\mathbf{\Gamma}\left(\nu + i\frac{\omega_0\beta}{2\pi}\right)\right|^2. \quad (10.53)$$

The shot/thermal noise crossover is recovered, in the tunneling limit:

$$S(\omega = 0) = 2e^*|\langle I_B\rangle|\coth(\omega_0\beta/2). \quad (10.54)$$

The above theoretical predictions have been verified in remarkable point contact experiments at filling factor $1/3$ in Saclay and at the Weizmann institute. Ref. [108] was performed at low temperatures in the shot noise dominated regime, while Ref. [109] used a fit to the thermal-shot noise crossover curve to identify the fractional charge. Subsequently, the Heiblum group also measured the fractional charge $e^* = e/5$ at filling factor $\nu = 2/5$ [110]. Experiments have also been performed in the strong backscattering regime, that is when the point contact splits the quantum Hall bar into two separate Luttinger liquids. Early reports suggested that the entities which tunnel are bare electrons – because they tunnel in a medium (vacuum) where Luttinger liquid collective excitations are absent – there is evidence that the noise deviates from the Poissonian noise of electrons. The noise at sufficiently low temperatures has been found to be super-Poissonian [111], with an effective charge $2e$ or $4e$ suggesting that electrons tunnel in bunches. There is no theoretical explanation of this phenomenon to this day.

On the theoretical side, an exact solution for both the current and the noise was found using the Bethe Ansatz solution of the boundary Sine-Gordon model [112]. It bridges the gap between the weak and the strong backscattering regimes. This work has also been extended to finite temperature [113], and careful comparison between theory and experiment has been motivated recently [114]. Noise at finite frequency has been computed in chiral Luttinger liquids using both perturbative techniques and using the exact solution at the (unphysical) filling factor $\nu = 1/2$ [106]. The noise displays a singularity at the "Josephson" frequency $e^*V/\hbar$.

A Hanbury-Brown and Twiss proposal has been made to detect the statistics of the edge state quasiparticle in the quantum Hall effect [115]. Indeed, the quasiparticle fields obey fractional statistics, and a noise correlation measurement necessarily provides information on statistics. The geometry consists of three edge states (one injector and two detectors) which can exchange quasiparticles by tunneling through the fractional Hall fluid. To leading order in the tunnel amplitudes, one finds the zero-frequency noise correlations:

$$\tilde{S}_{12}(0) = (e^{*2}|\omega_0|/\pi) T_1^r T_2^r R(\nu) \; , \tag{10.55}$$

where the renormalized transmission probabilities:

$$T_l^r = (\tau_0|\omega_0|)^{2\nu-2} \left[\tau_0 \Gamma_l / \hbar a\right]^2 / \Gamma(2\nu) \; , \tag{10.56}$$

correspond to that of a non interacting system at $\nu = 1$. Except for some universal constants, the function $R(\nu)$ can be measured experimentally by dividing the noise correlations by the product of the two tunneling currents. By comparing the noise correlations with those of an non-interacting system ($R(1) = -1$), one finds that the noise correlations remain negative at $\nu = 1/3$ although their absolute value is substantially reduced: there is less anti-bunching is a correlated electron HBT experiment. For lower filling factors, the noise correlations are found to be positive, but this remains a puzzle because perturbation theory is less controlled for $\nu < 1/3$.

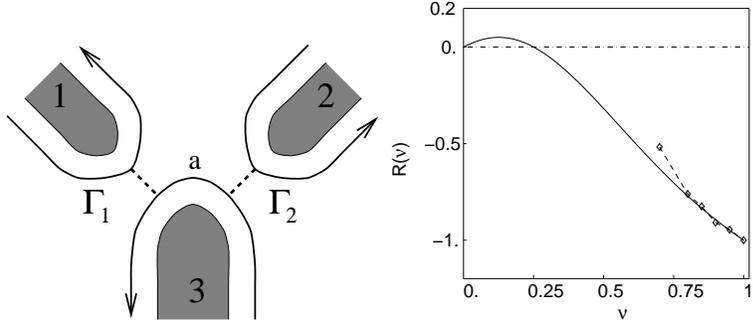

Fig. 23. Left: Hanbury-Brown and Twiss geometry with three edge states; quasiparticles are injected from 3 into 1 and 2 with tunnel hoppings $\Gamma_1$ and $\Gamma_2$. Right: normalized noise correlations as a function of filling factor; for comparison, the non-interacting value is $R(1) = -1$.

In the last few years, experiments studied the effect of depleting the edge state incoming on a point contact [116]. These experiments use a three edge state geometry as in Ref. [115]. Another point contact is in the path of this edge state in order to achieve dilution. If the former point contact is tuned as an opaque barrier, one normally expects poissonian noise from electrons. But because the edge state is dilute, quasiparticles seem to bunch together when tunneling. A noise diagnosis reveals that the effective charge predicted by Schottky's relation is in reality lower than the electron charge. These findings do not seem to find an appropriate theoretical explanation at this time [117].

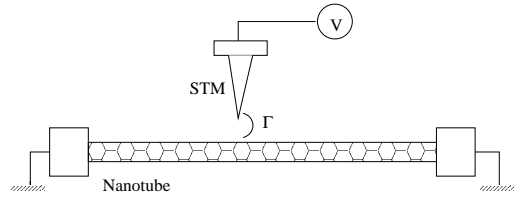

Fig. 24. Schematic configuration of the nanotube–STM device: electrons are injected from the tip at $x = 0$: current is measured at both nanotube ends, which are set to the ground.

### 10.6. *Effective charges in quantum wires*

As mentioned above, the chiral Luttinger liquids of the FQHE are an excellent test-bed for probing the role of interactions in noise. In a one dimensional quantum wire [98], the Luttinger liquid is non-chiral: interaction between right and left going electrons are effective over the whole length of the wire: the notion of a backscattering current is ambiguous. Nevertheless, non–chiral Luttinger liquids also have underlying chiral fields [118]. Such chiral fields correspond to excitations with anomalous (non-integer) charge, which has eluded detection so far. Here we mention briefly how shot noise measurements can provide information on such anomalous charges.

Carbon nanotubes, given the appropriate helicity, can have a metallic behavior. Due to their one-dimensional character, they are good candidates to probe Luttinger liquid behavior. Fig. 24 depicts a carbon nanotube with both ends grounded, but electrons are injected in the bulk of the tube with an STM tip. When electrons tunnel on the nanotube, they are not welcome because the excitations of this nanotube do not resemble electrons. This has been illustrated in tunneling density of states experiments [119]. The transport properties at the tunneling junction and in the nanotube can be computed using a Luttinger model for the nanotube, together with a perturbative treatment of the junction in the Keldysh formalism [120]. The tunneling electrons give rise to right and left moving quasiparticle excitations which carry charge $Q_+ = (1 + K_{c+})/2$ and $Q_- = (1 - K_{c+})/2$, where $K_{c+} < 1$ is the Luttinger liquid interaction parameter (in the absence of interactions or, equivalently, when the interactions in the nanotube are fully screened by a substrate, $K_{c+} = 1$). If the tunneling is purely local (say at $x = 0$), there is as much chance that $Q_+$ will propagate to the right while $Q_-$ propagates to the left than the opposite. The state of the outgoing quasiparticle excitations is entangled between these two configurations.

How can one identify the anomalous charges of quasiparticle excitations ? By performing a Hanbury-Brown and Twiss analysis of transport. Because bare electrons tunnel from the STM tip to the nanotube, the Schottky relation with charge $e$ holds for the tunneling current and the tunneling noise:

$$S_T \;\;=\;\; 2e\langle I_T\rangle \;. \tag{10.57}$$

However, non-integer charges are found when calculating the autocorrelation noise $S_\rho(x,x,\omega = 0)$ in one side (say, $x > 0$) of the nanotube together with the cross correlations $S_\rho(x,-x,\omega = 0)$ between the two sides of the nanotube:

$$S_{(}x,x,\omega = 0) \;\;=\;\; [1 + (K_{c+})^2]e\,|\langle I_\rho(x)\rangle| \sim (Q_+^2 + Q_-^2)\,, \tag{10.58}$$

$$S(x,-x,\omega = 0) \;\;=\;\; -[1 - (K_{c+})^2]e\,|\langle I_\rho(x)\rangle| \sim Q_+Q_- \;. \tag{10.59}$$

where $\langle I_\rho(x)\rangle$ is the average charge current at location $x$ in the nanotube. Note that the cross correlations of Eq. (10.59) are negative because we have chosen a different convention from Sec. 6: there the noise is measured away from the junction between the three probes. We conclude that the noise correlations are positive, which makes sense because excitations are propagating in both directions away from the junction.

The above considerations apply for an infinite nanotube, without a description of the contacts connected to the nanotube. It is known that the presence of such contacts (modeled by an inhomogeneous Luttinger liquid with $K_{c+} = 1$ in the contacts) leads to an absence of the renormalization of the transport properties [121] at zero frequency. Because of multiple reflections of the quasiparticles at the interface, the zero frequency noise cross correlations vanish at $\omega = 0$ due to the presence of leads. In order to retrieve information about anomalous charges, it is thus necessary to compute the finite frequency cross-correlations [122].

The result is depicted in Fig. 25. There are two competing time scales: $\tau_L = L/2v_{c+}$, the traveling time needed for Luttinger liquid excitations to reach the leads, and $\tau_V = \hbar/eV$, the time spread of the electron wave packet when the bias voltage between the tip and the nanotube is $V$. When $\tau_L \gg \tau_V$ (Fig. 25a), quasiparticle excitations undergo multiple bounces on the nanotube/contact interfaces, giving rise to multiple peaks in the autocorrelation noise. In the opposite case, the width of the tunneling electron wave packet is so large that there is only one maximum (Fig. 25b). For identifying anomalous charges, one should specify $\tau_V$, and the frequency should be tuned so that $\omega\tau_L$ is an odd multiple of $\pi/2$ (condition for a maximum), so that the sine function which enters. Then, one can measure experimentally the ratio $|S_{x,-x}/S_{x,x}| = (1-K_{c+}^4)/(1+K_{c+}^4)$ in order to extract the chiral charges $Q_\pm = (1 \pm K_{c+})/2$. Note that alternative finite frequency noise proposals to measure the anomalous charges in a Luttinger liquid with contacts have appeared recently in the literature [123].

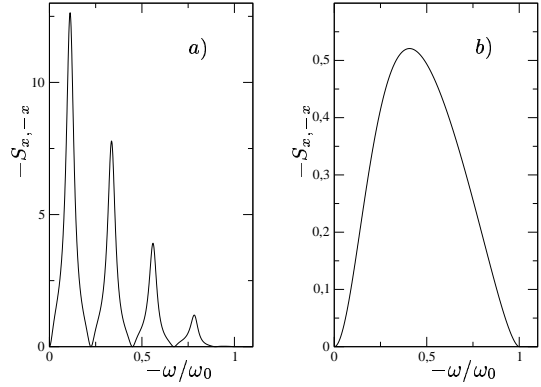

Fig. 25. Finite frequency noise correlations for a nanotube connected to leads. a) case where $\tau_L = 14\tau_V$; b) case where $\tau_L = \tau_V$.

## 11. Conclusions

The physics of noise in nanostructures is now one of the many exciting fields of mesoscopic physics. A motivation for this is the fact that compared to a current measurement, additional information can be extracted from the noise. Noise has been computed and measured experimentally in a variety of systems which are not described in this review. For additional approaches for calculating the noise, readers are directed towards articles and reviews. The few examples of noise calculations present in this course applied to situations where noise can either be characterized via scattering theory or, for the case of a Luttinger liquid, via perturbation theory in the tunneling Hamiltonian. Although this is a rather restrictive frame of work, these two situations can apply to a variety of nanodevices, possibly containing hybrid ferromagnetic or superconducting components.

Another generalization includes the discussion of noise in non-stationary situations, for instance when an AC bias is superposed to the DC voltage bias: photo-assisted shot noise was computed for a two terminal scatterer [124] and measured experimentally in diffusive metals [60], in normal metal superconducting junctions [69, 125], as well as in point contacts [126]. Photo-assisted shot noise provides an alternative way for measuring the effective charge of carriers. It has been recently computed in the context of the FQHE [127].

Noise measurements are typically hard measurements because one is dealing with a very small current or a small voltage signal, which needs to be "squared", and it is difficult to isolate the wanted noise from the unwanted one. The ratio of theory to experimental noise publications is still a bit too large. Yet experimental detection is making fast progress. Conventional noise apparatus, which convert a quantum measurement to a classical signal using for instance cold amplifiers continues to be improved. On the other hand, new measurement techniques use a noise detector which is part of the same chip as the device to be measured [56]. In such situations, it will be necessary to analyze what is the effect of the back-action of the measuring device on the circuit to be measured.

On the theoretical side, while the interest in computing noise remains at a high level, there is an ongoing effort to study the higher moments of the current, and the generating function of all irreducible moments. This sub-field bears the name of full counting statistics and was pioneered in the context of scattering theory in Refs. [128], but it is now generalized to a variety of systems. We refer the reader to the final chapters of Ref. [129]. Developments in this field include superconductor-hybrid systems as well as Luttinger liquids [130]. A nice seminar was presented during the les Houches school on this topic by A. Braggio, who applies full counting statistics to transport in quantum dots.

As discussed in the section on entanglement, more and more analogies can be found between nano-electronic and quantum optics, because both fields exploit the measurement of two-particle (or more) correlations. Bell inequality test allow to convince oneself that entanglement is at work in nano-devices. This has motivated several efforts to exploit this entanglement in a teleportation scenario. The entangler [90] can be used both as a generator of signet pairs as well as a detector of such pairs, and Ref. [131] describes a cell which teleports the state of an electron spin in a quantum dot to another electron in another quantum dot. The electron-hole entanglement scenario of Ref. [94] also gave rise to a teleportation proposal [132]. Interestingly, in order to control the output of such quantum information proposals, it is necessary to analyze many-particle correlations – or generalized noise – at the input and at the output of such devices.

Finally, I would like to emphasize that this course is the result of an ongoing effort over the years, and I wish to thank all my collaborators on noise since the early 1990's. Foremost, I should mention the role played by Rolf Landauer, who introduced me to noise. I am very much indebted to Gordey Lesovik, for his input and collaborations. Next I would like to thank my close associates in Marseille: Julien Torres who started his thesis working on NS junctions; Ines Safi, for her passage here working on the FQHE; my present collaborators Adeline Crepieux who provided latex notes on Luttinger liquids and who kindly read